# Security Patterns:
# A Systematic Mapping Study

Abbas Javan Jafari and Abbas Rasoolzadegan

**Abstract**— Security patterns are a means to encapsulate and communicate proven security solutions. They are well-established approaches for introducing security into the software development process. Our objective is to explore the research efforts on security patterns and discuss the current state of the art. This study will serve as a guideline for researchers, practitioners, and teachers interested in this field. We have conducted a systematic mapping study of relevant literature from 1997 until the end of 2017 and identified 403 relevant papers, 274 of which were selected for analysis based on quality criteria. This study derives a customized research strategy from established systematic approaches in the literature. We have utilized an exhaustive 3-tier search strategy to ensure a high degree of completeness during the study collection and used a test set to evaluate our search. The first 3 research questions address the demographics of security pattern research such as topic classification, trends, and distribution between academia and industry, along with prominent researchers and venues. The next 9 research questions focus on more in-depth analyses such as pattern presentation notations and classification criteria, pattern evaluation techniques, and pattern usage environments. The results and discussions of this study have significant implications for researchers, practitioners, and teachers in software engineering and information security.

## 1 INTRODUCTION

Software security patterns are structured solutions to reoccurring security problems (Uzunov et al., 2012), (Halkidis et al., 2008). These solutions encapsulate expert knowledge and best practices for developing secure software at different stages of the development lifecycle. From their initial introduction in 1997 (Yoder and Barcalow, 1997), security patterns have been steadily developed and integrated into software engineering and security methodologies (Uzunov et al., 2015), (Abramov et al., 2012a), (Nguyen et al., 2015). They are now among the well-established solutions for incorporating security into the software architecture and design (Rimba et al., 2015), (Alebrahim and Heisel, 2014b).

In the past two decades, there has been a considerable volume of research towards the field of software security patterns. Performing a comprehensive review of the resulting literature is crucial to identify the major trends and gaps in the field. It can serve as a reference guide for both new and experienced researchers interested in this field.

A comprehensive review in the entire research field of security patterns requires the use of a reliable, rigorous, impartial, and repeatable review technique. Systematic Literature Review (SLR) and Systematic Mapping Study (SMS) are two well-known methodologies in evidence-based software engineering to achieve these goals (Kitchenham et al., 2011), (Petersen et al., 2015). Even though the methodology for searching and extracting data in SLRs and SMSs are similar, these two types of systematic reviews have significant differences, particularly in their number of considered papers, research questions and data analysis techniques (Kitchenham et al., 2010), (Kitchenham et al., 2011), (Petersen et al., 2015). The research questions and analyses in systematic mapping studies are more general, while the number of covered papers are significantly larger, which corresponds to the goal of systematic mapping studies: identifying research trends and classifying topics which exist in a research field. This is different than systematic literature reviews which extract data from a small subset of research

_________________________

*A. J. Jafari*
*Faculty of Engineering, Ferdowsi University of Mashhad, Mashhad, Iran*

*A. Rasoolzadegan*
*Faculty of Engineering, Ferdowsi University of Mashhad, Mashhad, Iran*
*\* Corresponding author. E-mail: rasoolzadegan@um.ac.ir*

studies in an attempt to answer more specific research questions (Kitchenham et al., 2011). The goal of a systematic mapping study is not to analyze the detailed findings of each individual paper but rather, depict a coarse-grain mapping of relevant research into different classifications to answer corresponding research questions. A SMS is useful for providing general information on a topic and can be conducted as a pre-SLR task.

We have conducted a systematic review in the form of a mapping study on the relevant literature. A significant portion of the review period, and respectively, a significant portion of the paper is allocated to the search and data extraction methodology. We have included a comprehensive online supplementary document along with this paper (SupMat[1]) to present the details of our findings which were too large to fit in this paper. Throughout the paper, we will reference specific sections of this supplementary document for readers who are interested in the details of our methodology and the extracted data.

We have identified some other systematic mapping studies in the field, along with some surveys, so as to compare their methodology and results with our findings. The comparison reveals major differences between our work and previous reviews in terms of the number of papers studied, the considered scope of the studies, and the research methodology.

During the different stages of the search process, we found 8 secondary studies which review the security pattern literature. We have compared some of these works with our own study in Table 1. There were only two systematic reviews in our list ((Bunke et al., 2012), (Ito et al., 2015)), therefore we decided to include non-systematic reviews (classic surveys) in our comparison. This serves as a good opportunity to compare the attributes of a systematic process with traditional non-systematic methods. The surveys were selected based on publication venue and citation count.

TABLE 1

COMPARISON BETWEEN OUR WORK AND OTHER REVIEWS

| Ref. | Type | # of Studies | # of Venues | Search Strategy | # of DBs | Quality Criteria | Search Eval. | Time period | Scope |
|---|---|---|---|---|---|---|---|---|---|
| (Uzunov et al., 2012) | Survey | N/A | N/A | Database | 3 | ✘ | ✘ | 1997 to 2011* | Patterns and methodologies for securing distributed systems |
| (Yoshioka et al., 2008) | Survey | N/A | N/A | N/A | N/A | ✘ | ✘ | 1997 to 2007* | Published security patterns |
| (Bunke et al., 2012) | SMS | 67* | 11 | Manual* Snowballing Database | 2 | ✘ | ✘ | 1997 to mid-2012 | Studies that focus on presenting new patterns |
| (Ito et al., 2015) | SMS | 30 | N/A | Database | 2 | ✘ | ✘ | 2005 to Nov-2014 | Security pattern related research |
| **This Work** | **SMS** | **274** | **155** | **Manual Snowballing Database** | **4** | ✓ | ✓ | **1997 to the end of 2017** | **Security pattern related research (Development, Evaluation, Usage)** |

As can be seen in Table 1, our systematic review (last row) covers a substantially larger amount of studies and publication venues. This is mainly due to the utilization of an exhaustive 3-tier search strategy which is comprised of a manual search combined with a backward snowballing search, and a database search. In the manual search method, each venue in the search space (which is the current list of obtained venues-to-be-searched) is manually searched based on the keywords set. In backward snowballing, we aim to add new related papers by searching the reference list of currently added papers. Database search is a manual keywords-based search conducted on well-known databases such as IEEEXplore, ACM Library, ScienceDirect, and Springer Link. We have also performed an evaluation of our search strategy to ensure a high degree of completeness. This evaluation is further explained in section 2. This work aims to serve as a reference guide for security pattern research from 1997 to the end of 2017. Therefore, we have not limited the scope to a specific sub-field of security pattern research and instead, attempted to cover the entire research spectrum in this field. We initially planned to utilize only two search strategies (manual search and backward snowballing), but we found that these two approaches were lacking in a few specific instances. The issue with backward snowballing and manual search is that they rely on the cross-reference relationship between research papers. Papers referenced from multiple other papers, or published in commonly occurring venues are easily discovered, whereas isolated papers in unexpected venues have a chance of never being found. Another issue with backward snowballing stems from its backward nature. The search period for our study was up to the end of 2017, but to find 2017 papers through backward snowballing, we had to analyze papers published after that period (2018 and up) which was out of our scope. Similarly, newer papers with fewer citations are hard to discover through backward snowballing. Adding a database search proved beneficial as we discovered 97 new papers, 64 of which were included.

---

[1] http://sqlab.um.ac.ir/images/219/files/SQLab_SMS_SP-SupMat.pdf

The time period for the study of Uzunov et al. (Uzunov et al., 2012) and Yoshioka et al. (Yoshioka et al., 2008) in Table 1 are not explicitly stated, but are extracted based on the references list. In the work of Bunke et al. (Bunke et al., 2012), the mentioned number of studies consists of other non-primary studies (e.g. tech reports, books, surveys). This work utilizes three search strategies but the manual and snowballing searches are conducted separately (not combined) and the manual search is only conducted on a select set of conferences.

This paper is structured as follows. Section 2 focuses on presenting the research methodology used for this review. These include the 12 research questions, and the methods for searching, extracting, classifying, evaluating, and analyzing the data. We will also discuss the threats to the validity of our search and data extraction in this section. Section 3 presents the results of our systematic study categorized by the 12 research questions, along with the appropriate diagrams. We provide a discussion of our results in section 4. The implications stemmed from the results of this study for researchers, practitioners, and teachers interested in the field are presented in section 5. We conclude our work in section 6.

## 2 RESEARCH METHOD

The research methodology used in this paper is based on the updated guidelines for performing an SMS by Peterson et al. (Petersen et al., 2015) in which the authors have conducted an SMS on SMSs in the field of software engineering. Their set of updated guidelines stems from the lessons learned during their mapping study and also the guidelines presented by Kitchenham (Kitchenham, 2004), Arksey and O'Malley (Arksey and O'Malley, 2005), Biolchini (Biolchini et al., 2005), Kitchenham and Charters (Kitchenham and Charters, 2007), Peterson et al. (Petersen et al., 2008), and Budgen et al. (Budgen et al., 2008). The preliminary version of the presented methodology can be found in the work of Mayvan et al. (Mayvan et al., 2017) and also Ramaki and Rasoolzadegan (Ramaki et al., 2018). Fig. 1 depicts the two main phases of our review: the planning phase and the conducting phase. The transition from phase one to phase two is decided by the evaluation at the end of the planning phase. In this section, we will discuss the details of each step of our methodology.

### 2.1 Planning the Mapping Study

As can be seen in Fig. 1, the planning phase of our methodology is further decomposed into different stages. These stages include specifying the scope and research questions, the planning of the search process, the planning of the study selection process, specifying the search and study selection's evaluation strategy, and the planning of the data extraction and classification process. In the following, we will discuss each stage separately.

### 2.1.1 Specifying the scope and research questions

The foundation of a systematic study is the research questions (RQs). The effort to answer these questions guides the review process. In this study, we have defined 12 research questions and provided the rationality for each one (Table 2). These research questions are divided into two groups: primary and secondary. After answering the first three questions (primary), and based on the classification obtained by answering RQ2, we have further defined a set of 9 research questions (secondary) for a more in-depth analysis and discussion of each topic in the obtained classification.

In order to best answer the aforementioned research questions, the scope of this paper should cover all published works in the field. Our review covers a period of 21 years from 1997, the seminal paper introducing security patterns (Yoder and Barcalow, 1997), up to the end of 2017. We have ommited the year 2018 because this review was finished mid-2018 and including results from incomplete years would skew our trends and analyses. We have considered research works that provide an explicit contribution towards security patterns.

TABLE 2
THE RESEARCH QUESTIONS (RQS) FOR THIS SMS

| | # | Research Questions and Corresponding Rationale |
|---|---|---|
| **Primary Research Questions** | 1 | **How active is the field of security patterns and how is the research distributed between academia and the industry?**<br>Rationale: To identify the current volume of research and general trends in order to better depict the attractiveness of the field. Comparing the volume of research between academia and the industry can shed some light on the maturity of security of patterns in practical settings. |
| | 2 | **What are the core research topics in the field of security patterns and what is the publication trend and distribution for each topic?**<br>Rationale: To identify and classify the current research regarding security patterns, analyze the evolution and distribution of each topic and the potential trends in researcher's focus. |
| | 3 | **Which researchers and research venues are more active in this field and how is the research distributed geographically?**<br>Rationale: The demographics of security pattern research provide a useful starting point for interested researchers by identifying active scholars, venues, and countries. |
| **Secondary Research Questions** | 4 | **What standard notations and templates are used when presenting new patterns?**<br>Rationale: Using a suitable and established notation and template for presenting patterns can help pattern users grasp a more detailed, consistent and accurate understanding of the patterns. |
| | 5 | **How much attention has been given to pattern languages and systems in developing patterns?**<br>Rationale: Pattern are less effective in isolation and more effective when seen as a language of intertwined security solutions. It is important to analyze the extent that developers discussed how their pattern fits in the overall pattern landscape. |
| | 6 | **Which security objectives have the least/most corresponding security patterns? How much attention has been given to misuse/threat patterns?**<br>Rationale: Some security solutions might be more prominent than others, but pattern developers should take care to cover a varied spectrum of security solutions. |
| | 7 | **How have security patterns been classified?**<br>Rationale: General classification schemes might work to an extent, but a precise and comprehensive classification of security patterns should address security-specific criteria. |
| | 8 | **What techniques have been used to evaluate patterns?**<br>Rationale: Quantitative methods equip the evaluator with precise results whereas quantitative methods allow more freedom in evaluating more complex criteria. |
| | 9 | **Are evaluations generally applicable or are they pattern-specific? What is the consensus in regards to the usefulness of security patterns?**<br>Rationale: Evaluation strategies can be limited to specific types of patterns or cover the broad spectrum of current security patterns, with each method having its own benefits and drawbacks. Furthermore, pattern evaluations should shed some light on the overall usefulness of patterns as a security solution. |
| | 10 | **What environments have been investigated for using security patterns?**<br>Rationale: Using patterns in different environments might require specific considerations and therefore, justify separate research efforts for different environments. |
| | 11 | **Do pattern-based methodologies utilize all available patterns or are they pattern-specific?**<br>Rationale: There can be a trade-off in developing general purpose security methodologies which can utilize any patterns as an input, versus specific methodologies which only work with specific patterns. |
| | 12 | **What techniques have been used for selecting and applying security patterns?**<br>Rationale: Different selection and application techniques have been used in the security pattern literature. The maturity of these techniques will significantly impact the usability of security patterns. |

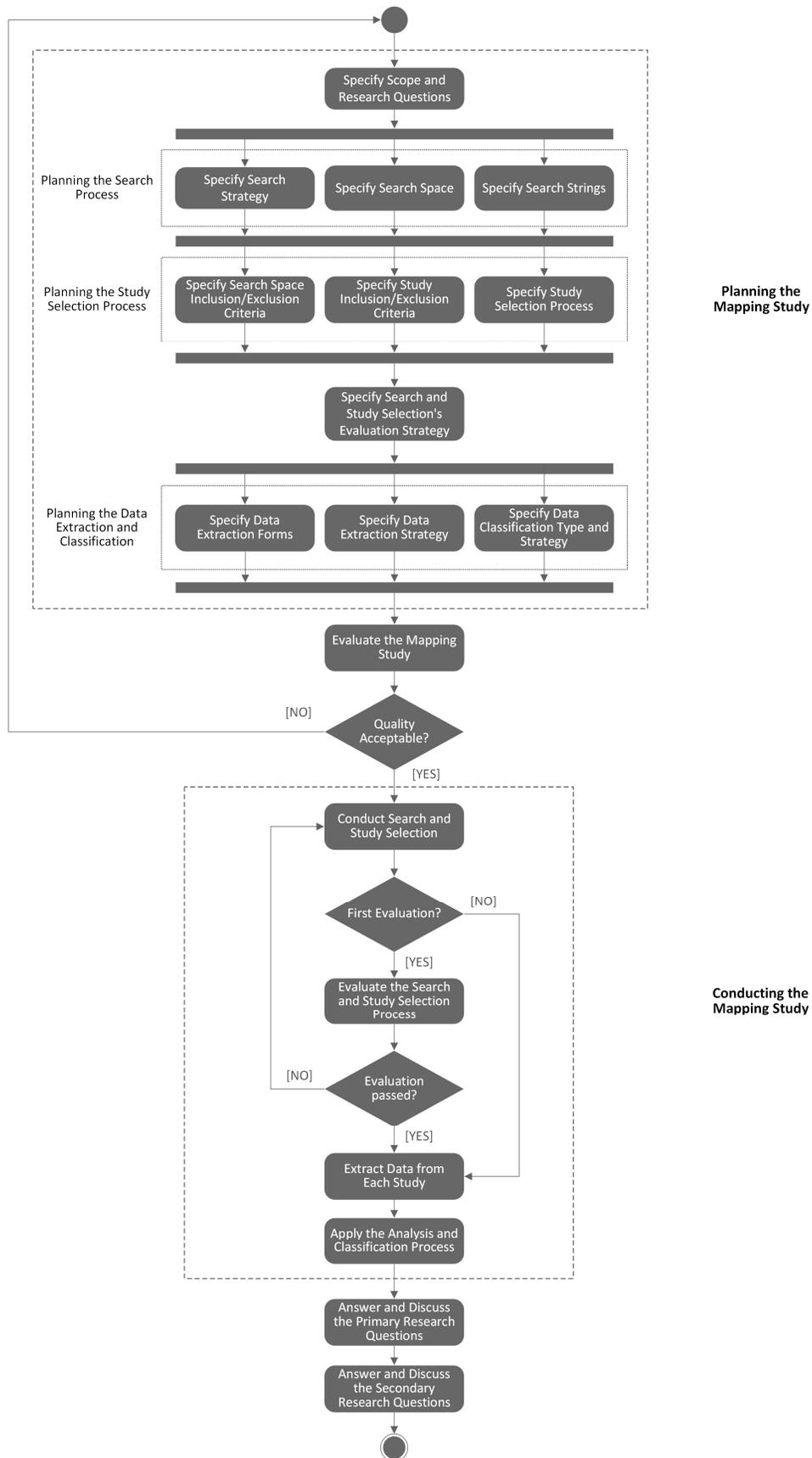

Fig. 1. The review methodology

### 2.1.2 Specifying the search strategy

When planning the search process, we must first specify the exact searching strategy which will be carried out during the review process. In this review, we have used three distinct search strategies. First, we combined backward snowballing and manual search to retrieve papers from 1997 to 2017. We then conducted a separate database search on the same time period. The decision to conduct a separate database search was to increase the overall number of retrieved papers, and also reveal possible shortcomings in the two previous strategies. This, in fact, resulted in a more complete set of retrieved papers, while revealing the inherent limitations of the backward snowballing and manual searches. Our overall search strategy is depicted in Fig. 2 and consists of seven steps. Identifying related journals, conferences, or workshops is not necessarily conducted after the manual search. Therefore, we have defined the S-Flag which is set to zero in step 2 and one in step 5.

1. The initial set of papers and keywords are extracted from a set of relevant secondary studies (by examining the cited papers in the secondary studies).
2. The corresponding search space (journal, conference, or workshop) of each included paper is extracted.
3. The set of extracted search spaces is evaluated based on a set of quality criteria and search spaces that satisfy these criteria are added. Manual search should be conducted on these search spaces using the pre-specified keywords set (which in this study is "Security pattern").
4. The study selection process is applied. If the paper is included, the process continues to step 5. If the paper is the result of backward snowballing, the process jumps to step 2.
5. Backward snowballing is conducted (on papers which have not been snowballed previously). This is carried out by examining the references of included papers to find new papers. If a new paper is found, the search process jumps to step 4. Otherwise, the process continues to step 6.
6. A database search is conducted using the pre-specified keywords set and new papers are extracted.
7. The study selection process is conducted on the newly found papers. The search process ends.

In order to conduct our database search, we have used the IEEE Xplore, ACM Digital Library, Springer Link, and ScienceDirect databases. However, many 2017 conference proceedings are published in early 2018. Therefore, the database search for the year 2017 was conducted twice; once in December 2017 and once in April 2018. This significantly improved the number of retrieved papers for 2017.

### 2.1.3 Specifying the search space

The search space of our process is initially empty. After extracting the initial set from the secondary studies, some journals, conferences, and workshops will be added to the search space. The search space will grow as the search progresses to include a list of relevant journals, conferences, and workshops. The complete set of journals, conferences, and workshops can be found in SupMat→Table 1.2 and Table 1.3.

### 2.1.4 Specifying the search strings

In our manual and database search, we used the keyword "security pattern" to retrieve all relevant papers and manually filtered the non-relevant results. The keyword was applied to the Title, Abstract, and Keywords section of the papers. Using more specific keywords is more time-efficient as it will reduce irrelevant results, but might also exclude some relevant papers. In this review, we opted for better completeness rather than time-efficiency.

### 2.1.5 Specifying the search space inclusion/exclusion criteria

When planning the study selection process, we must first specify the criteria to be used for including/excluding the search spaces. Every search space retrieved from the search process is initially included. The exclusion criteria in Table 3 are then applied to exclude some search spaces. We have used the JCR and SJR metrics for journals and Qualis, ERA and H5-Index metrics for conferences and workshops (metrics further explained in SupMat→Chapter 4).

The thresholds used for the exclusion criteria are obtained empirically. To achieve the appropriate thresholds, we considered two factors: 1) slight changes in the threshold values should not result in a major difference in the number of excluded/included papers and 2) applying the thresholds should not result in the exclusion of a lot of the highly cited papers.

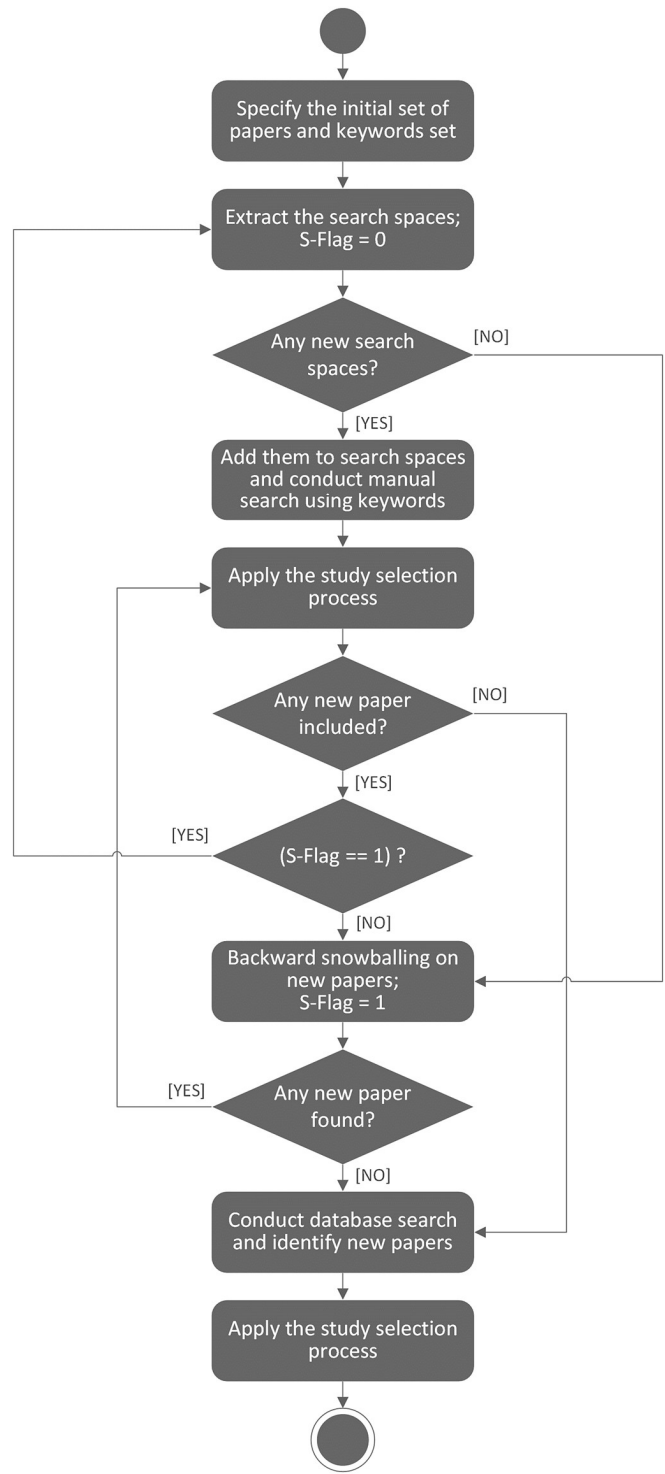

Fig. 2. The search strategy

### 2.1.6 Specifying the study inclusion/exclusion criteria

The inclusion criterion for the studies is the journal, conference, and workshop papers relevant to security pattern research. The term "relevant" is defined as a study with an explicitly stated (by the authors) contribution towards security pattern research. Papers with implicit contributions can hinder the objectivity of the results and are therefore not included. By using the search space exclusion criteria in the previous section, the manual search will not be conducted on excluded search spaces and therefore, papers belonging to those search spaces will not be retrieved. However, the backward snowballing and database search will retrieve and add relevant papers even if they belong to an excluded search space. This, along with other factors creates the need to define a set of exclusion criteria specific to the papers. These criteria are described in Table 4.

According to the criteria above, secondary studies (e.g. surveys, reviews) are excluded from our study set. However, these studies have been considered in Table 1 for comparison against our systematic study. The appropriate thresholds are defined based on the same factors mentioned in section 2.1.5.

TABLE 3

SEARCH SPACE EXCLUSION CRITERIA

| # | Description |
|---|---|
| 1 | If the journal is not indexed in JCR and SJR is below 0.4 |
| 2 | If the conference/workshop has a Qualis below B3 or ERA below B or H5-Index below 5 |
| 3 | If the journal/conference/workshop metrics are not available |

TABLE 4

PAPER EXCLUSION CRITERIA

| # | Description |
|---|---|
| 1 | The paper belongs to an excluded search space |
| 2 | The paper is published in a conference/workshop with a period of below 5 |
| 3 | The paper is not a primary study (e.g. survey) |
| 4 | Paper cannot be accessed (e.g. not indexed) |

Even though these criteria are empirically defined to ensure that the majority of high-quality works are included, applying them causes a small subset of highly-cited papers to be excluded. Therefore, we calculated the average citations of the included papers in each year and compared them with the set of excluded studies in those years. Excluded studies which have a citation count of 2 times the average (or more) are included, regardless of other factors. Detailed statistics can be found in SupMat→Table 1.4 and Table 1.5.

### 2.1.7 Specifying the study selection strategy

The study selection strategy process depicted in Fig. 3 consists of two main parts. In the first part, we evaluate the paper's relevance by reading the title, abstract, and keywords (and the full-text when necessary). If the paper is irrelevant, it is discarded, and if the relevance is uncertain (due to a disagreement between reviewers), a third-party should judge its relevance. Infrequent disagreements during the study selection process were resolved following a discussion among Software Quality Lab (SQL) members. In the second part of the process, we evaluate the paper using the criteria in Table 4.

### 2.1.8 Specifying the search and study selection evaluation strategy

In order to evaluate the search strategy, we can choose between subjective evaluation (conducted by expert review) and objective evaluation (through quantitative criteria). To evaluate our search and study selection, we have used the quasi-sensitivity metric to provide a more objective evaluation (rather than just using an expert's opinion). Sensitivity (Altman and Bland, 1994) is a well-known evaluation technique (borrowed from medicine) to evaluate the quality and efficiency of a search strategy (Zhang et al., 2011). It is calculated using the equation below (Equation 1):

$$Sensitivity = \frac{\# \: of \: studies \: in \: our \: SMS}{\# \: of \: studies \: overall} \times 100 \qquad (1)$$

However, as we can never know the exact number of overall studies, we have used the quasi-gold standard (QGS) metric (Zhang et al., 2011). QGS refers to a set of studies from well-known sources in that research community. To create this gold standard, we received help from members of the Software Quality Lab. The lab members referred to the webpages of the pioneer researchers in the field and extracted their relevant papers.

This set of papers, after applying the inclusion/exclusion criteria, constitutes our QGS.

We have used the QGS to calculate the quasi-sensitivity and compare the result with a predefined threshold. If the result falls below the threshold, we should redo the search and study selection process using the QGS. According to Zhang et al. (Zhang et al., 2011), an acceptable threshold falls between 70% and 80%. We have used 80% as our predefined threshold. As will be mentioned later in section 2.3.2, the actual result of the evaluation for this work yields 95.5%, which is much higher than our pre-defined threshold.

### 2.1.9 Planning the data extraction and classification process

In order to extract the data from each study during our search, we prepared a set of data extraction forms. Each field in these data forms should be mapped to one of the RQs. We have shown a sample of our data field categorization into two separate forms in Table 5.

The "Topic" field is derived from the "Keywords" field using a keywords clustering algorithm. The resulting research tree for the topics is explained in section 3.

TABLE 5
ITEMS OF THE DATA EXTRACTION FORM FOR RQ1-RQ3

| Form# | Item | Rationale |
|---|---|---|
| 1 | Title of study | Identifying the papers |
| 2 | Authors name | For answering RQ3 |
| 2 | Country | For answering RQ3 |
| 1 | Keywords | For answering RQ2 |
| 1 | Topic | For answering RQ2 |
| 2 | Publication Year | For answering RQ1 & RQ2 |
| 2 | Venue | For answering RQ3 |

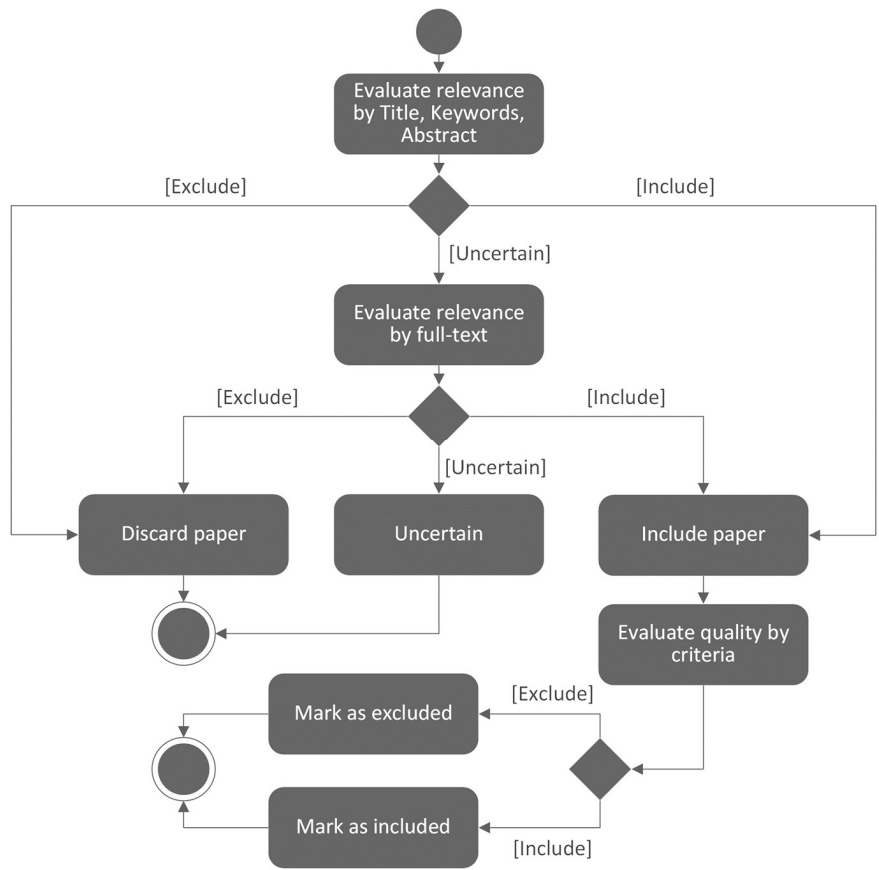

Fig. 3. The study selection strategy

## 2.2 Evaluating the Mapping Study

For the evaluation of our mapping study, we have used the evaluation rubrics from the work of Peterson et al. (Petersen et al., 2015). Table 6 presents the rubrics, the actions we took toward each rubric, and our corresponding evaluation degrees (no description, minimal, partial, or full) based on the guidelines in the work of Peterson et al. (Petersen et al., 2015). The last column of the table is taken from Peterson et al. (Petersen et al., 2015) which shows the percentage of each evaluation degree for the mapping studies examined by their research team. Based on the evaluation degrees of the rubrics in this study compared to other studies, we believe our approach has reached an acceptable level of quality and we can continue on to the next phase (conducting the mapping study). If for any reason, the evaluation degrees were not acceptable, the planning phase should be revised before continuing to the next phase.

## 2.3 Conducting the Mapping Study

After specifying the search and study selection process in the planning phase of the mapping study, and after achieving suitable evaluation measures, we can conduct our mapping study. This phase is also comprised of multiple stages. These include conducting the search and study selection, evaluating the search and study selection, data extraction, and analysis and classification. This section will clarify the activities conducted in each stage.

### 2.3.1 Conducting the search and study selection process

The first step in starting the search process involves an initial set of papers which are extracted from a set of secondary studies. These secondary studies are found using an informal search process by using the keywords "security pattern" along with other keywords such as "survey" and "review". After we retrieved a set of secondary studies (SupMat→Table 1.1), we extracted relevant cited papers from those studies. These papers constitute our initial set (SupMat→Tables 1.6 and 1.7). The manual and backward snowballing search are then carried out on the included studies in the initial set. After the manual and backward snowballing search is finished, a database search is carried out to complement the results set. Table 7 and Table 8 provide sample extraction tables for journal and conference/workshop search spaces and Table 9 presents the statistics for the search space set. The terms "JEC" and "CEC" refer to the venue exclusion criteria in Table 3. Table 10 presents a sample extraction table for journal papers and Table 11 presents the statistics for the studies set. The term "PEC" refers to the paper exclusion criteria in Table 4. In Table 11, the years before 2006 were omitted from the journal papers section, as these years only included conference or workshop publications.

TABLE 6
EVALUATION RUBRICS

| Rubric | Actions | Evaluation degree | Evaluation degree of the mapping studies examined by (Petersen et al., 2015) |
|---|---|---|---|
| Specifying the scope and research questions | The study motivations and the research questions are provided. | Partial | 100% partial |
| Specifying the search strategy | All three search strategies (Manual, Backward snowballing, Database) have been utilized. | Full | About 58% no description, about 23% minimal, about 19% full |
| Evaluating the search and study selection process | The quasi-gold standard has been used as a test set for evaluation. Disagreements between reviewers are resolved following a discussion with lab members. Decision rules are defined for study inclusion/exclusion. | Partial | About 44% no description, about 42% minimal, about 14% partial |
| Data extraction and classification | Disagreements between reviewers are resolved following a discussion with lab members. The included studies are classified based on type, venue, and topic. | Minimal | About 10% no description, about 61% minimal, about 29% partial |
| Threats to validity | Threats to validity have been discussed. | Full | About 13% no description, about 87% full |

TABLE 7
SAMPLE EXTRACTION TABLE FOR JOURNAL SEARCH SPACES

| Journal Name | IF | SJR | Q | Publisher | Exclusion Criteria |
|---|---|---|---|---|---|
| IEEE Transactions on Dependable and Secure Computing | 1.59 | 1.16 | 1 | IEEE | - |
| Computers & Security | 1.64 | 1.02 | 1 | Elsevier | - |
| Innovations in Systems and Software Engineering | - | 0.2 | 4 | Springer | JEC1 |
| International Journal on Advances in Security | - | - | - | IARIA | JEC2 |

TABLE 8
SAMPLE EXTRACTION TABLE FOR CONFERENCE AND WORKSHOP SEARCH SPACES

| Conference Name | H5-Index | Qualis | Core (ERA) | Library | Exclusion Criteria |
|---|---|---|---|---|---|
| Conference on Pattern Languages of Programs (PLoP) | 10 | B3 | B | ACM | - |
| IEEE International Conference on Web Services (ICWS) | 25 | A1 | A | IEEE | - |
| Canadian Conference on Electrical and Computer Engineering (CCECE) | 15 | 0 | C | IEEE | CEC1 |
| International Conference on Engineering secure software and systems (ESSoS) | - | - | - | Springer | CEC2 |

### 2.3.2 Evaluating the search and study selection process

The lab members prepared a set of 134 papers by examining the webpages of pioneer researchers. After removing the duplicates (due to multiple author papers) and applying the inclusion/exclusion criteria, we acquired a QGS of 67 papers, 3 of which did not exist in our study set. Therefore, we achieved a sensitivity of 95.5% which is well above our predefined threshold. More details on the final evaluation set (QGS) can be found in SupMat→Table 2.1.

Even though the evaluation result exceeds the required threshold, it is interesting to discuss why these 3 papers were missed. By taking a closer look, we found that the papers were either new and lacking significant citations (unable to discover using backward snowballing search), the venues were not in our selected databases (unable to discover using database search), and/or the papers were "isolated" occurrences in unrelated venues which do not regularly publish such papers (unable to discover using manual search on the search space).

### 2.3.3 Data extraction

In this step, we extracted paper data based on the data extraction forms. Disagreements during the data extraction process were resolved following a discussion with members of the Software Quality Lab. These

TABLE 9
SEARCH SPACE STATISTICS

| Search Space | Initial Set | | Manual & Snowballing | | Database Search | | Total |
|---|---|---|---|---|---|---|---|
| | Included | Excluded | Included | Excluded | Included | Excluded | |
| Journals | 5 | 2 | 4 | 5 | 9 | 2 | 27 |
| Conferences/Workshops | 15 | 19 | 13 | 22 | 24 | 35 | 128 |
| Total | 20 | 21 | 17 | 27 | 33 | 37 | 155 |

TABLE 10

SAMPLE EXTRACTION TABLE FOR JOURNAL PAPERS

| Paper Title | Journal | Year | Exclusion Criteria |
|---|---|---|---|
| Security patterns modeling and formalization for pattern-based development of secure software systems | Innovations in Systems and Software Engineering | 2016 | PEC1 |
| Securing distributed systems using patterns: A survey | Computers & Security | 2016 | PEC2 |
| Security Patterns for Untraceable Secret Handshakes with optional Revocation | International Journal on Advances in Security | 2010 | PEC1 |
| Architectural Risk Analysis of Software Systems Based on Security Patterns | IEEE Transactions on Dependable and Secure Computing | 2008 | - |

disagreements were infrequent (less than 10 cases) and were mostly related to the main goal/contribution of a selected paper, especially in cases where the authors mention multiple contributions. Due to the relatively large amount of data, the results of the data extraction step are presented in SupMat→Tables 3.4 to 3.14.

### 2.3.4 Data analysis and classification

The data extracted in the previous step is analyzed and documented based on the aforementioned RQs. We have presented the data for the topic classification in SupMat→Tables 3.4 to 3.5 which correspond to RQ1 and RQ2. The results of this step are further described in section 3.

One of the main objectives of this SMS is to classify the research topics regarding security patterns. The classification is achieved using a two-part technique which involves keyword clustering and expert's analysis. First, we apply a keyword clustering technique, as mentioned by (Febrero et al., 2014) and (Noyons et al., 2000), on the keywords of the included set of papers. The paper's keywords are comprised of the author keywords and the manually extracted keywords from the abstract (which represent the main contribution of the paper). Authors have many goals for their papers and relying solely on author keywords is problematic. In order to extract the main goal(s) of the paper, we can refer to the abstract where the authors generally provide an explicit explanation of the goals and contributions of their paper using commonly used keywords such as "defining patterns", "categorization" and "methodology" (if not, we must refer to the full text). To accomplish this, we first extracted the 15 most frequent keywords (keywords with an occurrence rate of 10 or above). These keywords can be found in SupMat→Table 3.3. We then applied a hierarchical clustering algorithm with complete linkage on the keyword normalized co-occurrence matrix (Noyons et al., 2000). The co-occurrence matrix depicts the number of times each two keywords appeared together (SupMat→Chapter 3). We then normalized the co-occurrence matrix based on the cosine index (Equation 2).

$$sim(i.j) = \frac{\sum_{k=1}^{n} X_{ik} X_{kj}}{\sqrt{((\sum_{k=1}^{n} X_{ik}^2)(\sum_{k=1}^{n} X_{kj}^2))}} \quad (2)$$

where $X_{ij}$ is the number of co-occurrences of the keyword i with keyword j.

In the recalculated matrix, the similarity of keywords will be based on the cognitive orientation of two keywords in relation to all other keywords. The keyword clustering process results in 5 clusters. Following an expert's analysis among the authors and the Software Quality Lab members, these 5 clusters are merged (where possible) into 3 final clusters based on their cognitive meaning. The expert intervention is necessary because the classification of research topics should take the cognitive meaning of the keywords into account and not be solely based on keywords co-occurrence. However, the clustering process guides the expert into determining the number of topic categories in the final classification.

TABLE 11
PAPER STATISTICS

| Search Space | Year | Initial Set | | Manual & Snowballing | | Database Search | | Total |
|---|---|---|---|---|---|---|---|---|
| | | Included | Excluded | Included | Excluded | Included | Excluded | |
| Journals | 2017 | 0 | 0 | 2 | 1 | 5 | 2 | 10 |
| | 2016 | 0 | 0 | 0 | 1 | 4 | 1 | 6 |
| | 2015 | 0 | 0 | 3 | 0 | 0 | 0 | 3 |
| | 2014 | 0 | 0 | 4 | 0 | 1 | 0 | 5 |
| | 2013 | 0 | 0 | 1 | 0 | 1 | 1 | 3 |
| | 2012 | 0 | 0 | 2 | 3 | 0 | 0 | 5 |
| | 2011 | 0 | 0 | 1 | 0 | 0 | 0 | 1 |
| | 2010 | 0 | 1 | 1 | 1 | 0 | 0 | 3 |
| | 2009 | 0 | 1 | 2 | 1 | 0 | 0 | 4 |
| | 2008 | 1 | 0 | 2 | 1 | 0 | 0 | 4 |
| | 2007 | 1 | 0 | 0 | 0 | 0 | 0 | 1 |
| | 2006 | 3 | 0 | 0 | 1 | 0 | 0 | 4 |
| **Sub-Total (Journals)** | **2006-2017** | **5** | **2** | **18** | **9** | **11** | **4** | **49** |
| Conferences/Workshops | 2017 | 0 | 0 | 6 | 0 | 3 | 5 | 14 |
| | 2016 | 0 | 0 | 7 | 0 | 7 | 7 | 21 |
| | 2015 | 0 | 0 | 12 | 1 | 4 | 9 | 26 |
| | 2014 | 0 | 0 | 21 | 4 | 3 | 4 | 32 |
| | 2013 | 0 | 0 | 13 | 2 | 1 | 4 | 20 |
| | 2012 | 0 | 1 | 10 | 8 | 2 | 5 | 26 |
| | 2011 | 1 | 4 | 10 | 3 | 4 | 5 | 27 |
| | 2010 | 4 | 2 | 9 | 4 | 1 | 3 | 23 |
| | 2009 | 7 | 3 | 9 | 10 | 0 | 3 | 32 |
| | 2008 | 9 | 5 | 9 | 5 | 0 | 1 | 29 |
| | 2007 | 5 | 8 | 9 | 6 | 1 | 2 | 31 |
| | 2006 | 6 | 7 | 4 | 6 | 0 | 0 | 23 |
| | 2005 | 4 | 1 | 1 | 1 | 1 | 1 | 9 |
| | 2004 | 6 | 1 | 2 | 1 | 0 | 0 | 10 |
| | 2003 | 8 | 2 | 1 | 1 | 0 | 0 | 12 |
| | 2002 | 7 | 0 | 1 | 1 | 0 | 0 | 9 |
| | 2001 | 4 | 1 | 0 | 0 | 0 | 0 | 5 |
| | 2000 | 2 | 0 | 0 | 0 | 0 | 0 | 2 |
| | 1999 | 1 | 0 | 0 | 0 | 0 | 0 | 1 |
| | 1998 | 1 | 0 | 0 | 0 | 0 | 0 | 1 |
| | 1997 | 0 | 1 | 0 | 0 | 0 | 0 | 1 |
| **Sub-Total (Conference/Workshop)** | **1997-2017** | **65** | **36** | **124** | **53** | **27** | **49** | **354** |
| **Total (Overall)** | **-** | **70** | **38** | **142** | **62** | **38** | **53** | **403** |

TABLE 12
THE KEYWORDS AND DESCRIPTION OF EACH CATEGORY

| Category | Keywords | Description |
|---|---|---|
| Pattern Development | -Pattern name (e.g. Secure logger)<br>-Presenting/Providing/Describing patterns<br>-Classification/Categorization<br>-Pattern language/catalog<br>-Enhanced/Improved patterns<br>-Specifying patterns/Pattern template | The research focused on the development of the patterns themselves. This includes presenting new patterns or pattern languages, new pattern templates, enhanced versions of previous patterns, and classification schemes. |
| Quality Evaluation | -Quality evaluation/analysis<br>-Pattern evaluation/analysis | The research focused on assessing the quality of security patterns or the impact of applying security patterns on software quality. |
| Pattern Usage | -Using/Employing/Utilizing patterns<br>-Security engineering/Secure software development<br>-Pattern-based methods<br>-Applying/Application of patterns | The research focusing on using security patterns to improve other methods, incorporating security patterns with software development methodologies, or techniques for applying patterns. |

The final classification of topics is depicted in Table 12. We mapped each paper to its corresponding topic based on its keywords. Papers belonging to multiple categories were discussed and mapped to the single most suitable category. The small subset of papers (5%) which could not be mapped to any topic was mapped to the "Miscellaneous" category. Table 13 presents a sample extraction table we used for the keywords and categories.

## 2.4 Threats to the validity of the search and data extraction process

One of the important evaluation rubrics mentioned in Table 6 is the discussion of threats to validity. During the SMS process, various factors during the planning phase or the conducting phase can threaten the validity of the results. In the following, we will discuss the major threats and corresponding countermeasures.

One of the issues faced by systematic reviews is the completeness of the collected studies. In this SMS, we have utilized all three major search strategies presented in the literature (manual search, snowballing search and database search) whereas most systematic studies utilize only two search strategies (Petersen et al., 2015). In order to further ensure the reliability of the search spaces and the final included studies set, we also conducted an evaluation of the search and study selection phase, which yielded above-satisfactory results. These considerations reasonably ensure the completeness of our search and the resulting studies set.

TABLE 13
SAMPLE EXTRACTION TABLE FOR PAPER'S KEYWORDS AND CATEGORY

| Paper Title | Keywords (Extracted keywords) | Category |
|---|---|---|
| Enterprise security pattern: A new type of security pattern | Secure information system, Enterprise security architecture, Security pattern, Enterprise security pattern, Threat modeling (New type of security pattern) | Development |
| Automated verification of security pattern compositions | Automated verification of security pattern compositions (Automated verification) | Evaluation |
| ASE: A comprehensive pattern-driven security methodology for distributed systems | Secure software engineering, Security methodologies, Distributed systems security, Security patterns, Threat patterns, Security solution frames (Presenting a pattern-driven security methodology) | Usage |
| How to integrate legal requirements into a requirements engineering methodology for the development of security and privacy patterns | Security and privacy patterns, Legal requirements, organization, Pattern validation, Healthcare | Miscellaneous |

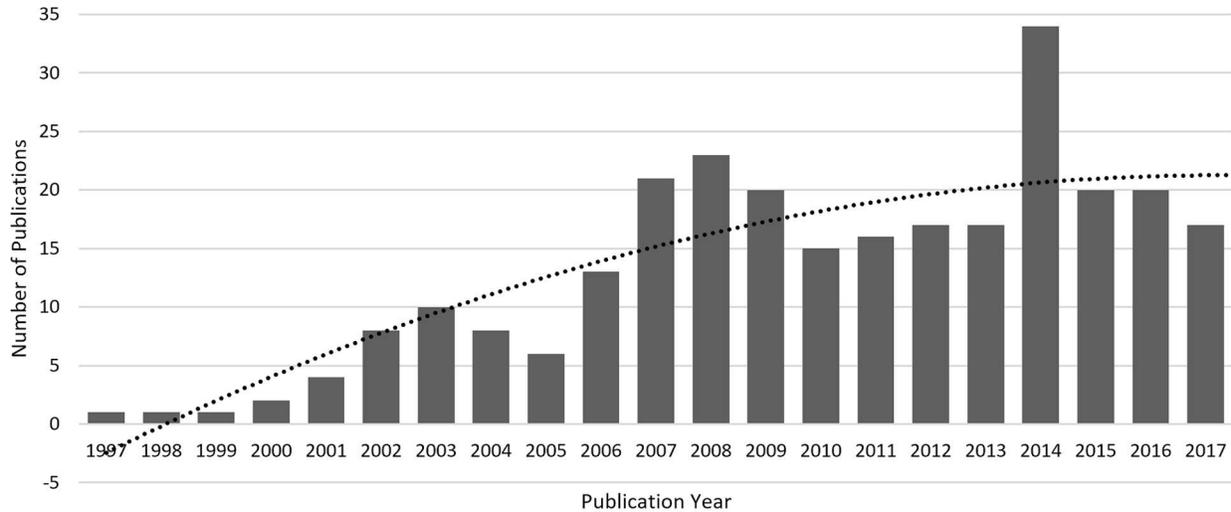

Fig. 4. Frequency of publications per year

The issue of reviewer bias and misunderstanding can arise during the study selection process. To address this issue, the study selection process was reviewed by members of the Software Quality Lab. Furthermore, the evaluation was conducted by separate members of the Software Quality Lab to minimize the probability of intentional and unintentional biases.

Another threat arises from the process of data extraction and classification. As mentioned earlier, a significant number of studies either lack any author keywords or the keywords provided are not specific enough to internally classify security patterns research. In order to supplement these shortcomings, we have extracted keywords from the abstract (and full-text if necessary) which accurately depict the contribution of the paper. These extracted keywords are checked by a secondary reviewer to remedy potential misunderstandings. In choosing the names of each category for the classification of topics, we opted for the most inclusive name after a discussion among the lab members.

## 3 RESULTS OF THE STUDY

We have categorized the results of our systematic mapping study according to the research questions defined at the beginning of the review process (Table 2). The answers to each research question are based on the extracted data. The analyses of the results are discussed extensively in section 4. The results in this section are strictly based on the included studies set (based on inclusion/exclusion criteria). We have used appropriate diagrams to best represent the extracted data. RQ1-RQ3 depict the demographics of security patterns including trends, distributions and topic classification. RQ4-RQ7 address the development topic, RQ8-RQ9 relate to the evaluation topic, and RQ10-RQ12 focus on the usage topic.

### 3.1 The frequency of publications and distribution between academia and industry (RQ1)

Observing the frequency of publications depicts how much attention the research field has received over time and whether the field is currently trending or not. Fig. 4 displays the frequency of publications from the year 1997 (where the field was first introduced) to the end of 2017. Observing the trend-line (polynomial with a degree of 2) for this 21-year period shows an overall upward trend for the publication frequency. The diagram in Fig. 4 reveals some anomalies in the publication trend, particularly in the year 2014. These are partly caused by special issues in journals or specially-focused workshops. For instance, the journal of "Computer Standards and Interfaces" has a special issue "Security in Information Systems: Advances and new Challenges" at June 2014, or the focus of the "2014 IEEE Symposium on Service-Oriented System Engineering" is on "Software Engineering Methods". Apart from these issues, we haven't observed any other reasons for these anomalies.

An interesting observation is the distribution of research between academia and industry, depicted in Fig. 5. The term "industry" represents university-independent research labs, consulting firms, and other industries pursuing research towards the field. As can be seen in Fig. 5, the majority of authors are affiliated with academia. More complete statistics for the distribution between academia and industry for 39 countries can be found in SupMat→Table 3.8.

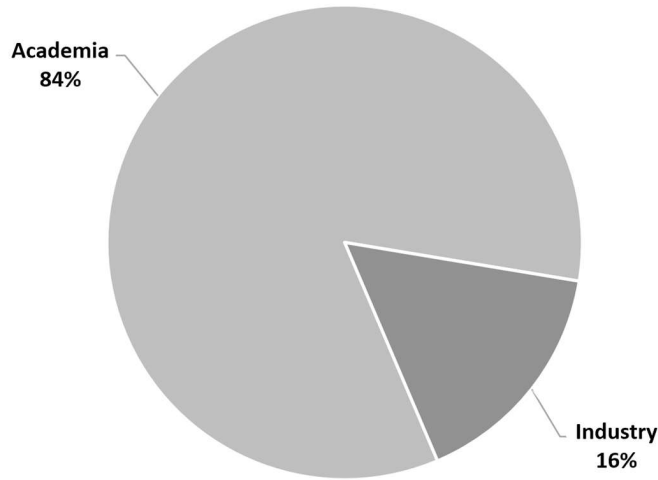

Fig. 5. Distribution of research between academia and industry

### 3.2 Core research topics in the field of security patterns along with trends and distributions (RQ2)

In section 2.3, we used a keyword clustering technique along with expert's analysis to classify the research in the field of security patterns. As a result, the classification tree in Fig. 6 shows the main topics (categories) and sub-topics (sub-categories) in security patterns research. The *miscellaneous issues* category represents a small subset of papers which could not be classified under any of the main topics. The pie chart in Fig. 7 presents an overview of the research load in each category. In the following, we will elaborate each of the aforementioned categories.

### 3.2.1 The Pattern Development category

This category includes research towards the development of the patterns themselves. As can be seen in Fig. 7, this category makes up 52% of the security pattern research field and has received the most attention between the three categories. This category includes:

- Introducing a new pattern or pattern specification structure: Many research works focus on presenting new patterns corresponding to new solutions. Examples include security patterns at the requirements phase (Mazo and Feltus, 2016), (Alebrahim and Heisel, 2014a), the architecture design phase (Syed and Fernandez, 2015), (Moral-García et al., 2014), the detailed design phase (Steinegger et al., 2016), (Fernández et al., 2014), or the test phase (Smith and Williams, 2012). Some research works have focused on the development of more suitable templates and specification structures for presenting and documenting patterns (Menzel et al., 2010), (Hamid and Percebois, 2014).
- Presenting a new form of security patterns or enhancing other patterns with security features: Some research works have focused on introducing a new form of security patterns such as misuse/threat patterns (Alkazimi and Fernandez, 2016), (Sulatycki and Fernandez, 2015), or enterprise security patterns (Moral-García et al., 2014), (Moral-García et al., 2014). Some other works such as (Jafari and Rasoolzadegan, 2016) and (Fernandez and Ortega-Arjona, 2009) have focused on the idea of adding security capabilities to previously un-secure design patterns.
- Pattern languages and the Classification (or Categorization) of patterns: Developments in this regard include attempts at classifying security patterns (Anand et al., 2014), (Fernandez et al., 2008) or presenting new pattern languages (Hafiz et al., 2012), (Li et al., 2013), (Mundie et al., 2012). It's worth noting that a pattern language is not just a catalog of patterns, but also a system to guide the designers in using multiple patterns to design secure systems (Hafiz, 2013).

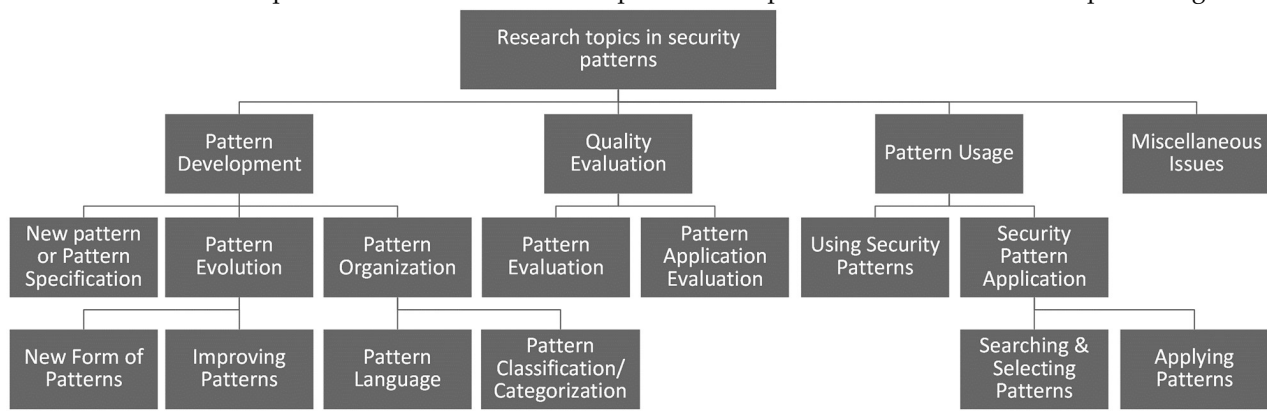

Fig. 6. Security patterns' research tree

The most frequent keywords in the pattern development category are displayed in Table 14.

TABLE 14
TOP KEYWORDS OF THE "DEVELOPMENT" CATEGORY

| Keyword(s) | Frequency |
|---|---|
| Pattern Name(s) | 50 |
| Presenting Pattern(s) <br> Describing Pattern(s) <br> Proposing Pattern(s) | 41 |
| Requirements Pattern(s) <br> Requirements Engineering | 21 |
| Security Architecture <br> Architecture Pattern(s) | 20 |
| Pattern Language <br> Pattern Catalog | 19 |
| Misuse Patterns <br> Threat Patterns | 17 |
| Pattern Classification <br> Categorization | 15 |

### 3.2.2 The Quality Evaluation category

This category includes research which quantitatively and qualitatively evaluate security patterns. As can be seen in Fig. 7, this category makes up 8% of the security pattern research field and has received the least attention amongst the three categories. This category includes:
- Evaluating security patterns: Some research works such as (Halkidis et al., 2006), (Motii et al., 2016b) and (Duncan and de Muijnck-Hughes, 2014) have focused on the evaluation of security patterns themselves.
- Evaluating the effect of security patterns on software systems: Another group of research works such as (Abramov et al., 2012b), (Smith and Williams, 2012) and (Ortiz et al., 2011) have analyzed the effectiveness of security patterns and pattern-based methods on the security of the target system.

The most frequent keywords in the quality evaluation category are displayed in Table 15.

TABLE 15
TOP KEYWORDS OF THE "EVALUATION" CATEGORY

| Keyword(s) | Frequency |
|---|---|
| Pattern Evaluation | 15 |
| Pattern Validation <br> Pattern Verification | 7 |
| Assurance | 5 |
| Testing <br> Model-based Testing | 4 |

### 3.2.3 The Pattern Usage category

This category includes research towards the utilization of security patterns in secure software development. As can be seen in Fig. 7, this category makes up 35% of the security pattern research field. This category includes:

- Using security patterns: Many research works such as (Uzunov et al., 2015), (Abramov et al., 2012a) and (Nguyen et al., 2015) focus on integrating security patterns with software development and engineering methodologies to create new approaches or to improve pre-existing methods.
- Security pattern application: Another group of research works have focused on the techniques relating to the selection and application of appropriate security patterns using manual or automatic approaches (Guan et al., 2016), (Motii et al., 2016a), (Bouaziz and Kammoun, 2015).

The most frequent keywords in the pattern usage category are displayed in Table 16.

TABLE 16
TOP KEYWORDS OF THE "USAGE" CATEGORY

| Keyword(s) | Frequency |
|---|---|
| Security Engineering Security Methodologies Pattern-based Development | 30 |
| Using (Misuse) Patterns Employing Patterns | 18 |
| Selecting Patterns | 10 |
| Applying Patterns | 6 |

### 3.2.4 The publication trend and distribution for each core topic

One of the important aspects of our analysis was to observe the publication frequency of each category independently. The independent trends of each category are displayed in SupMat→Chapter 3. The data has been aggregated into the line graph of Fig. 8. This presents the evolution of each category as well as a comparison between the various categories. The pie charts in Fig. 9 depict the distribution of each topic into its internal subtopics.

Interestingly, the publication trend of the development category has the closest resemblance to the overall trend of security pattern publications. The bulk of current research in the development category belongs to presenting new patterns. The quality evaluation category has received the least attention in the past two decades. However, as with the other two topics, this topic has been gaining increased attention in the recent years. The distribution of research in the evaluation category is split almost evenly between the evaluation of the patterns themselves, and the evaluation of the impact of security patterns on software quality. The pattern usage category is following a similar trend as the pattern development category, albeit with some reasonable delay. The current distribution of this category is skewed in favor of research towards integrating security patterns with software development and engineering methodologies to create new approaches or to improve pre-existing methods. Research towards different techniques for selecting and applying security patterns have received less attention.

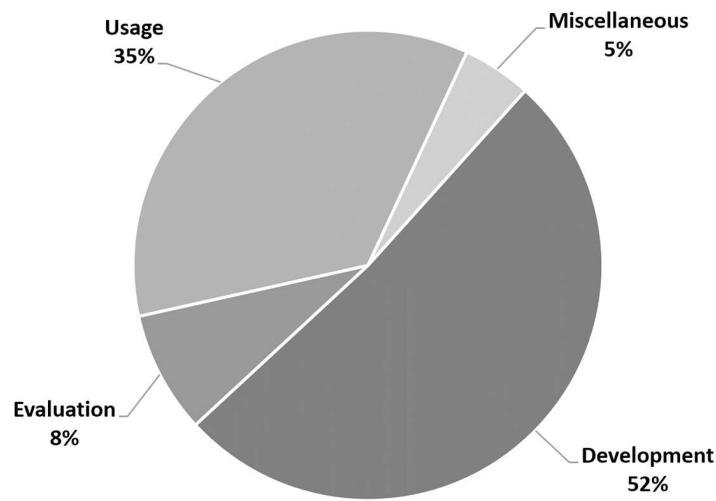

Fig. 7. Distribution of topics in the literature

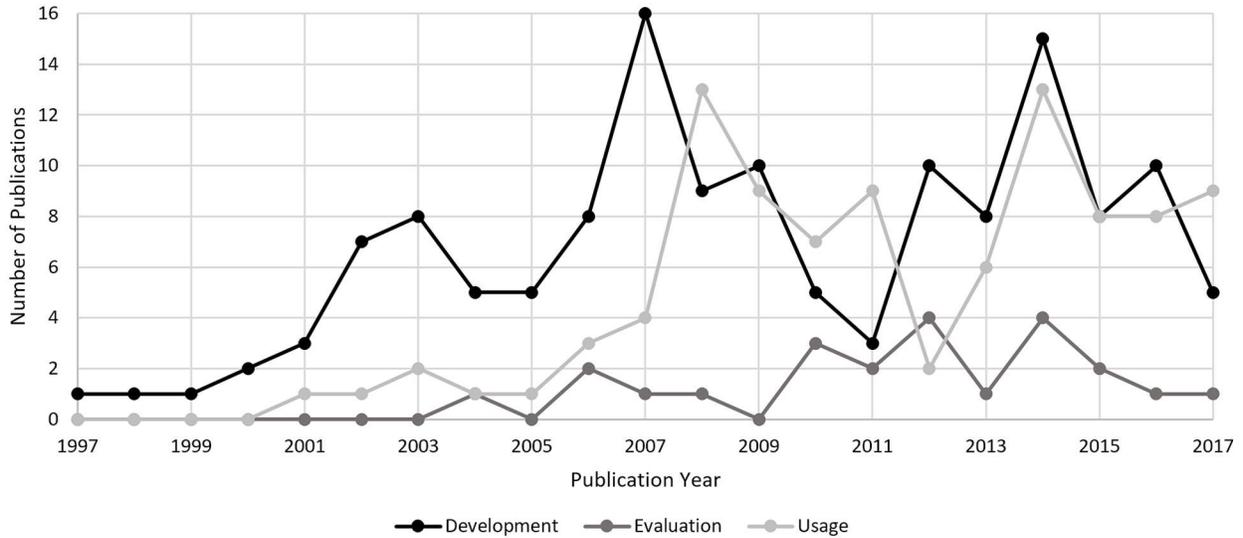

Fig. 8. Comparison of the publication frequency for each core topic

### 3.3 Active researchers, research venues and geographic distribution (RQ3)

Identifying key researchers in the field of security patterns is beneficial to those who wish to enter this field. Fig. 10 displays the top researchers (with publication count of over 10). The researchers are sorted from left to right based on the decreasing order of publications. Eduardo B. Fernandez leads the list with 79 publications, Followed by Nobukazu Yoshioka, Maria M. Larrondo-Petrie, Hironori Washizaki, Munawar Hafiz, Antonio Mana, and Brahim Hamid. The size of bubbles in each category represent the number of publications in each topic. Even though some researchers such as A. Mana have focused more on the Usage topic, the general research distribution in the top researchers depict a similar distribution as in Fig. 7.

The most active Conferences/Workshops in the field of security patterns are depicted in Fig. 11. The "International Conference on Pattern Languages of Programs" and the "European Conference on Pattern Languages of Programs" are the premier venues for pattern research and are consequently on top of the list.

TABLE 17

SAMPLE EXTRACTION TABLE FOR ANSWERING RQ3

| Paper Title | Author(s) | Affiliation | Journal |
|---|---|---|---|
| Enterprise security pattern: A new type of security pattern | Santiago Moral- García, Santiago Moral- Rubio, David G. Rosado, Eduardo B. Fernández, Eduardo Fernández-Medina | Rey Juan Carlos Univ. (Spain) BBVA Group (Spain) Univ. of Castilla-La Mancha (Spain) x2 Florida Atlantic (USA) | Security and Communication Networks |
| Automated verification of security pattern compositions | Jing Dong, Tu Peng, Yajing Zhao | UT Dallas (USA) x3 | Information and Software Technology |
| Building a security reference architecture for cloud systems | Eduardo B. Fernandez, Raul Monge, Keiko Hashizume | Florida Atlantic (USA) Adelaide (Australia) Universidad Técnica Federico Santa María (Chile) | Requirements Engineering |

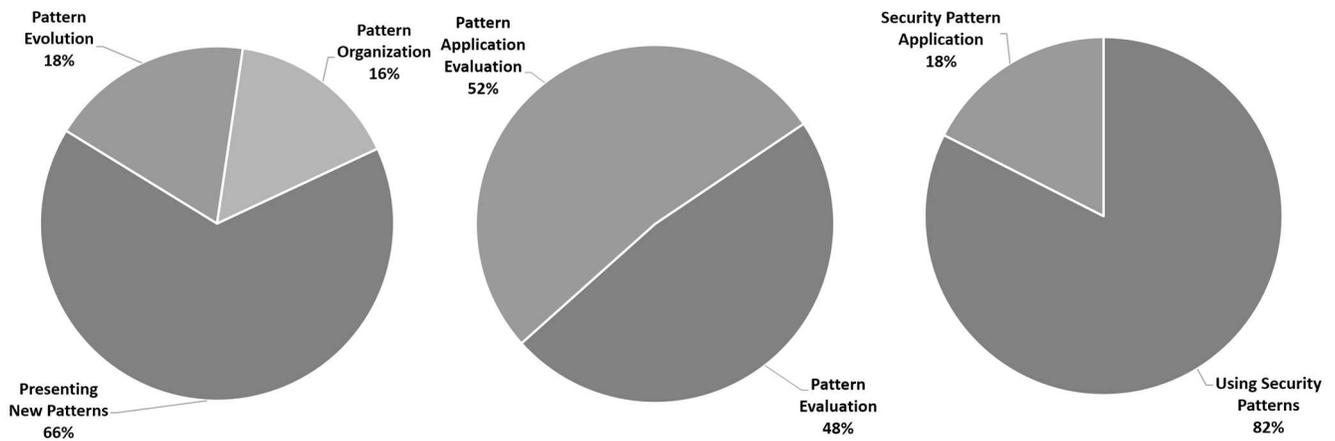

Fig. 9. Distribution of sub-topics in the literature

Others include the "International Workshop on Database and Expert Systems Applications", the "International Conference on Availability, Reliability and Security", the "International Conference on Advanced Information Systems Engineering", and the "International Requirements Engineering Conference". Conferences and Workshops with a publication count of below 5 are not shown in the figure (but are included in the studies set). Also, some conferences such as the "AsianPLoP" and "VikingPLoP" are not on the list because they were previously excluded based on the inclusion/exclusion criteria for search spaces.

The geographical distribution of publications helps researchers identify the pioneer institutions conducting security patterns research around the globe. The diagram in Fig. 12 displays the result of 274 included papers from 1997 to the end of 2017. The statistics are based on the affiliations of the papers. Many papers have multiple authors, therefore, we have considered the country per author, not per paper, resulting in a total of 791 affiliations. Table 17 presents a sample extraction table for data regarding RQ3.

As can be seen in Fig. 12, USA is leading by 31%, followed by Germany, Japan, Spain, France and the UK. Unsurprisingly, all these countries reside in the top 10 of the SCImago Country Rank (SCImago, 2015). Countries with an affiliation count of less than 20 have been grouped in the "Others" category.

## 3.4 Standard notations and templates for presenting new patterns (RQ4)

We have investigated the degree to which security patterns have utilized standard notations and templates. From the 141 papers in the development category, 109 papers have focused on presenting new patterns or an improved version of a previously identified pattern. The rest, mostly belonging to the pattern organization sub-category, do not qualify for this discussion and they have been omitted. The statistics relevant to this research question are displayed in Table 18. The complete list of papers, along with the respective collected data, can be found in SupMat→Tables 3.9 and 3.10.

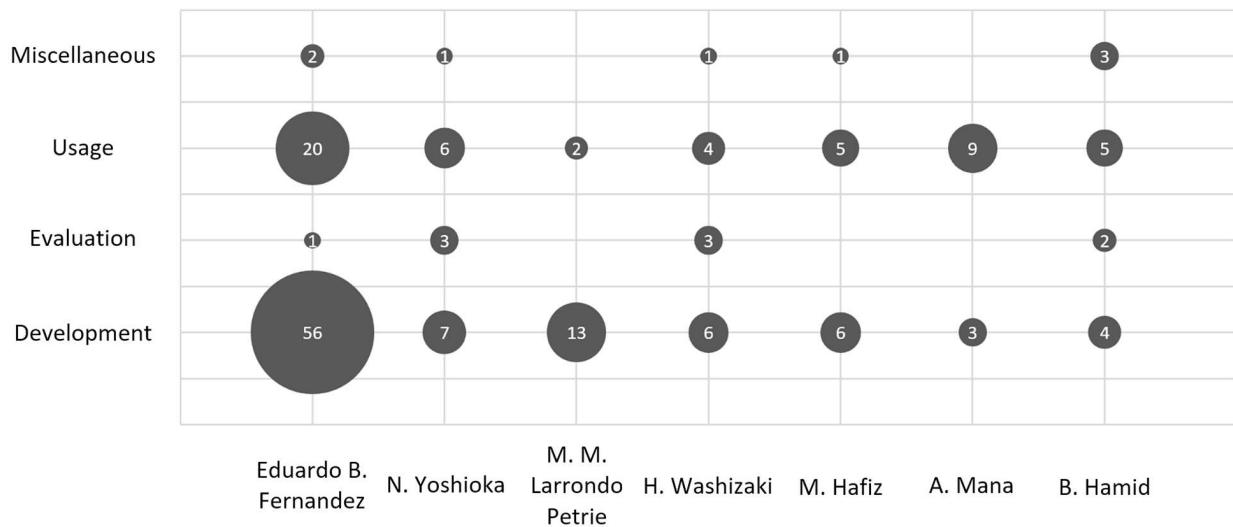

Fig. 10. Top researchers in the field of security patterns

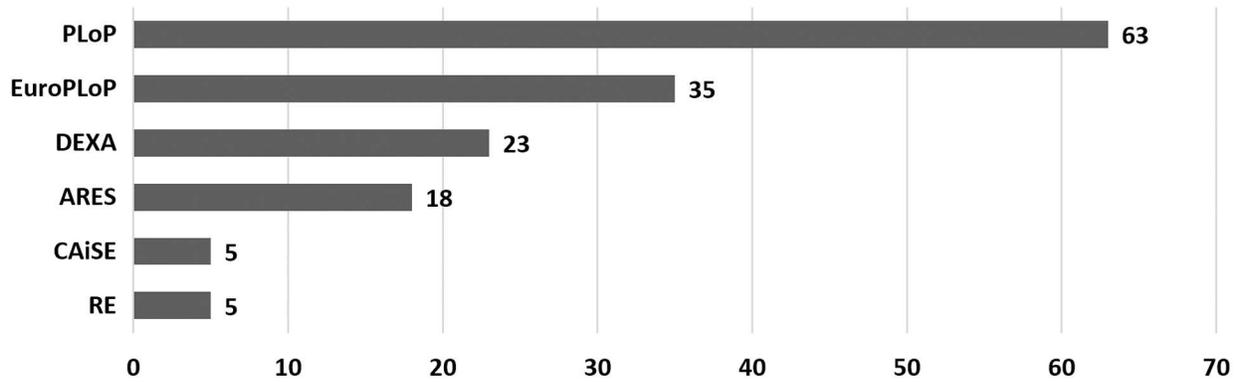

Fig. 11. Top conferences in the field of security patterns

When presenting new security patterns, authors seem to prefer UML as their top choice, with 69% of studies utilizing the UML class diagram and sequence diagram to present the structure and behavior of their patterns. We have also observed other notations such as BPMN (Ahmed and Matulevičius, 2014) and Tropos (Mouratidis et al., 2005), but the sum of these secondary notations amount to less than 5% of the studies we investigated. A rather significant observation is that 27% of papers do not use any standard notation to present their patterns.

The most common template for describing patterns is the POSA template which adds up to 64% of the papers. The relatively popular GoF template (Gamma et al., 1995) has achieved less popularity among security pattern authors, making up only 8% of the templates used in the studies. Interestingly, for 22% of the papers, the authors have opted for a custom (often security-centric) pattern template. This doesn't mean that the authors have devised completely new templates from scratch; rather, many have added security-related sections to previously established templates such as the ones in POSA or GoF. An example is adding a section to express the "security forces" associated with each pattern (Sinnhofer et al., 2016). We observed that 6% of papers have not used any template to present their patterns.

### 3.5 Importance of pattern languages when developing new patterns (RQ5)

We have investigated the degree to which security pattern research works have made efforts to discuss related patterns. As stated previously in RQ4, in order to answer this research question, we only consider papers that have focused on presenting new patterns or an improved version of a previously identified pattern. The rest of the papers in the development category, mostly belonging to the pattern organization sub-category, have been omitted. We can observe in Table 18, that the vast majority of papers presenting new patterns (83%) have taken care to mention, and sometimes discuss, related patterns. The complete list of papers, along with the respective collected data, can be found in SupMat→Tables 3.9 and 3.10.

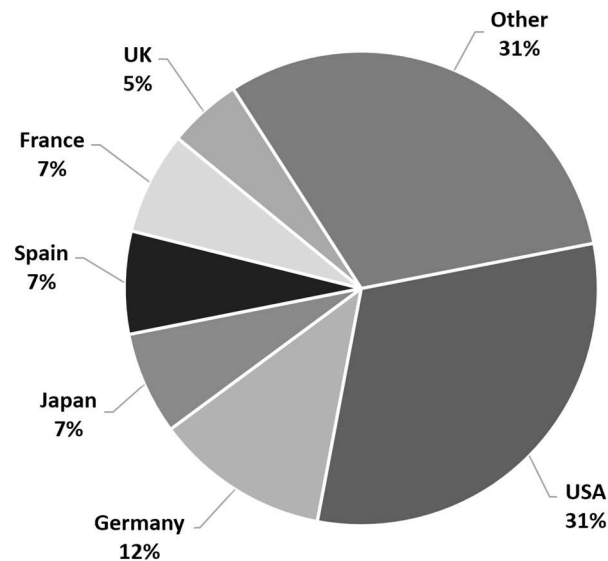

Fig. 12. Geographical distribution of researchers

### 3.6 The most and least sought after security objectives for pattern developers and the emergence of misuse patterns (RQ6)

Security isn't a singular solution, rather, it's a collection of many different objectives and their corresponding solution categories. These range from authentication and authorization to administration and monitoring. Our findings reveal a significant tendency towards specific solution categories. This was first observed by a study in 2007 (Heyman et al., 2007) in which the authors mentioned the lack of attention to objectives such as recovery and privacy. Out of the 141 papers we investigated in the development category, 44 have focused on specific objectives. A significant number of papers focused on multiple objectives which is why the sum of percentages in table 18 for the security objectives surpass 100%. Among these objectives, authentication and authorization respectively appear in 36% and 50% of the studies. On the other end of the spectrum, monitoring and privacy respectively appear in 9% and 14% of the studies. Although authentication is also used to ensure integrity, it is often grouped with authorization to compose access control solutions. We reported their results separately as grouping authentication and authorization together would further exacerbate the imbalance in the results. The complete list of papers, along with the respective collected data, can be found in SupMat→Tables 3.9 and 3.10.

Misuse patterns are an emerging concept which was first introduced in (Fernandez et al., 2009). From a total of 92 papers that present new patterns, 8% are devoted to misuse patterns. These types of patterns attempt to view the system from an attacker's viewpoint in order to identify potential targets and techniques to circumvent security measures. They are a means to model the recurring steps and involved components which attackers utilize to exploit software systems. In this sense, misuse patterns can be considered as the opposite of a security pattern, but they are nonetheless a part of the security pattern research community due to their tightly coupled concepts and their reference to security patterns (to explain what patterns can mitigate each part of the attack). It should be noted that misuse patterns are different than anti-patterns (Brown et al., 1998) which present common bad practices that should be avoided by developers. Papers discussing misuse patterns, along with the respective collected data, can be found in SupMat→Tables 3.4 and 3.5.

### 3.7 Classification schemes for organizing security patterns (RQ7)

Several different classification schemes have been proposed in the literature (Hafiz et al., 2007), (Fernandez et al., 2008), (Washizaki et al., 2009). We have investigated the various classification schemes and criteria used in security patterns literature. We have identified 9 papers which present a classification scheme for security patterns. We have then compiled common classification criteria along with their sample values in Table 19. The criteria used are listed next to the relevant papers in SupMat→Tables 3.9 and 3.10. As shown in Table 19, patterns can be classified based on a variety of facets which can belong to the field of information security or software engineering. Thus, we can use criteria from both software engineering (e.g. development lifecycle) and information security (e.g. security objective) to classify patterns. A significant number of papers have utilized multiple criteria which is why the sum of percentages in table 18 for the classification criteria surpass 100%.

### 3.8 Techniques for evaluating security patterns (RQ8)

Evaluation attempts are necessary to assess the degree to which security patterns satisfy their objective. We have investigated the 23 papers in the evaluation category to identify common approaches and techniques. Our observations reveal that 24% of papers have used quantitative methods in security pattern evaluation, as opposed to the 71% of papers which have opted for qualitative methods (the remaining 5% have focused on issues such as discussing pattern specification techniques which are irrelevant to this comparison). Quantitative evaluation can be further categorized into concrete techniques. For example, the authors in (Dong et al., 2010) have formally defined the behavioral aspects of security patterns and used model checking to automatically verify security pattern compositions. On the other hand, the work in (Kobashi et al., 2013) utilizes model testing to verify the proper application of security patterns. The same can be said for qualitative evaluation, which consists of a variety of techniques. The two most common methods are experts' analysis and empirical studies. The complete list of papers for this analysis, along with the respective collected data, can be found in SupMat→Tables 3.11 and 3.12.

TABLE 18
STATISTICS FOR RQ4-RQ7

|  | Type | # of Studies | % of Studies |
|---|---|---|---|
| Notation | UML | 75 | 69% |
|  | BPMN | 2 | 2% |
|  | Tropos | 2 | 2% |
|  | None | 29 | 27% |
| Template | POSA | 70 | 64% |
|  | GoF | 8 | 8% |
|  | Custom | 24 | 22% |
|  | None | 6 | 6% |
| Related Patterns | Considered | 90 | 83% |
|  | Not Considered | 19 | 17% |
| Security Objective | Authentication | 16 | 36% |
|  | Authorization | 22 | 50% |
|  | Privacy | 6 | 14% |
|  | Monitoring | 4 | 9% |
|  | Others | 3 | 7% |
| Classification Criteria | Concern | 3 | 33 |
|  | Boundary | 3 | 33 |
|  | Lifecycle | 4 | 44 |
|  | Others | 8 | 89% |

### 3.9 The focus of pattern evaluations and the consensus regarding pattern usefulness (RQ9)

Our literature review reveals that not all evaluation efforts are focused on a similar type of security patterns. Based on our observations, 39% of studies in the evaluation category have focused their evaluation efforts towards a specific pattern or set of patterns. For example, the work in (Halkidis et al., 2006) focuses on the patterns presented by Blakley et al. (Blakley and Heath, 2004), and the work in (Smith and Williams, 2012) focuses on security test patterns. Evaluation efforts should ultimately provide feedback on the usefulness of security patterns and provide guidance for future improvement. Therefore, after investigating the techniques used for evaluation, and the specific patterns used by those techniques, it is time to analyze the evaluation results. The majority of empirical evaluations have concluded the overall usefulness of security patterns. According to Abramov et al. (Abramov et al., 2012b), using a pattern-based method for secure development results in higher security and faster completion time, and it is also easier and clearer to use. The evaluation of security test patterns by Smith and Williams (Smith and Williams, 2012) concludes that using these security patterns can help novices generate similar black-box tests as experts, implying that patterns are indeed effective in disseminating expert knowledge among novice designers. There are also some studies which discuss their reservations about the usefulness of security patterns. These studies don't refute the usefulness of security patterns completely, rather, they discuss the lack of evidence for some of the positive claims made about patterns or they believe that the current state of security patterns needs some improvement before it can satisfy those claims. For example, Yskout et al. (Yskout et al., 2015) believe that there is not enough evidence to support the claims about security patterns leading to more secure designs. They also point to the sub-optimal quality of the documentation for current security patterns as a possible explanation. A similar criticism to the current state of security pattern documentation has been expressed by the work of Bunke (Bunke, 2015), in which the author discusses the lack of concrete code examples in security patterns. Based on our observations, the majority of studies in the evaluation category are convinced of the usefulness of security patters for creating more secure designs and reducing the number of sub-optimal design decisions in the process. Others are mostly concerned about some of the current shortcomings which need to be addressed before claims about the usefulness of patterns can be fully satisfied. The complete list of papers for this analysis, along with the respective collected data, can be found in SupMat→Tables 3.11 and 3.12.

TABLE 19

SECURITY PATTERN CLASSIFICATION CRITERIA

| Classification Criteria | Sample Values | Examples |
|---|---|---|
| Objective | Confidentiality, Integrity, Availability (CIA). | (Ponde et al., 2016), (Hafiz et al., 2007) |
| Lifecycle | Requirements, Architecture Design, Detailed Design, Implementation, Test. | (Ponde et al., 2016) |
| Architecture | Application, DBMS, Network, OS, Distribution. | (Washizaki et al., 2009), (Fernandez et al., 2008) |
| Goal | Prevention, Detection, Response. | (Mundie et al., 2012), (Washizaki et al., 2009) |
| Domain | Enterprise, Firewall, Real-Time, Medical… | (Washizaki et al., 2009) |
| Boundary | Core, Perimeter, Exterior. | (Ponde et al., 2016), (Hafiz et al., 2007) |
| Threats | Microsoft's STRIDE classification for threats. | (Ponde et al., 2016), (Hafiz et al., 2007) |
| Vulnerability | OWASP's vulnerability classification. | (Anand et al., 2014) |

### 3.10 Environments for investigating pattern usage (RQ10)

An important indicator to determine the maturity and usefulness of security patterns is the volume and maturity of research towards their usage in software systems. A significant portion of research towards pattern usage is aimed at specific environments. The variety in usage environments also depicts the wide range of applications for security patterns. Among the papers in the usage category, 56% have not bounded their work to a specific environment and instead, provided methodologies and application techniques which are independent of the environment. However, the remaining 44% have conducted their research in specific settings. Our statistics, based on the results of the SMS, indicate that distributed and cloud architectures are respectively responsible for 6% and 8% of research efforts in the usage category. There are also multiple studies which have focused on the usage of security patterns in environments such as embedded systems (Hamid et al., 2013), software defined networking (Petroulakis et al., 2016), web applications (Rosado et al., 2006) and database (Abramov et al., 2012a). The complete list of papers for this analysis, along with the respective collected data, can be found in SupMat→Tables 3.13 and 3.14.

### 3.11 Generality or specificity of pattern-based methodologies (RQ11)

In section 3.9, we discussed the specificity of pattern evaluation research towards specific patterns. There is a similar discussion in regards to the focus of pattern-based methodologies on a specific group of patterns. Based on our observations, 42% of the studies in the usage category have focused on a specific set of security patterns. These can be further broken down into patterns for security and dependability (S&D), authentication, authorization, threat, analysis, requirements, VoIP, cloud, and privacy. The complete list of papers for this analysis, along with the respective collected data, can be found in SupMat→Tables 3.13 and 3.14.

### 3.12 Techniques for selecting and applying security patterns (RQ12)

Methodologies for integrating security patterns into the software development lifecycle often focus on coarse-grain concepts and outlines the required steps in a relatively abstract manner. In order to guide software practitioners into the proper utilization of security patterns in their projects, the security pattern research community must attend to the fine-grain details of selecting and applying patterns in practice. We have observed a variety of pattern selection techniques in the literature including the utilization of feature models (Nguyen et al., 2015), (Slavin et al., 2014), ontology (Arjona et al., 2014), (Guan et al., 2014), and selection rules (Li et al., 2014), (Pearson and Shen, 2010). There is a lesser degree of variety in the techniques used for applying patterns. Almost every paper we have observed utilizes model transformation techniques to apply patterns in the design or code. The complete list of papers used for this analysis, along with the respective collected data, can be found in SupMat→Table 3.13 and Table 3.14. We have presented more detailed statistics for RQ10 to RQ12 in Table 20.

TABLE 20
STATISTICS FOR RQ10-RQ12

|  | Type | # of Studies | % of Studies |
|---|---|---|---|
| **Usage Environment** | Generic | 43 | 56% |
|  | Distributed | 5 | 6% |
|  | Cloud | 6 | 8% |
|  | Others | 23 | 30% |
| **Pattern Types** | Generic | 45 | 58% |
|  | Cloud | 2 | 3% |
|  | S&D | 10 | 13% |
|  | Threat | 2 | 3% |
|  | Others | 18 | 23% |
| **Selection Method** | Ontology | 4 | 23% |
|  | Alternative Specification | 3 | 18% |
|  | Selection Rules | 3 | 18% |
|  | Others | 7 | 41% |
| **Application Method** | Model Transformation | 13 | 93% |
|  | Decision Support Map | 1 | 7% |

## 4 DISCUSSION OF THE RESULTS

This section includes our analyses and discussions regarding the results in section 3. Instead of presenting a separate discussion for each of the 9 research questions, we have categorized our discussion into 4 subsections which collectively covers the entire results of this study. This deliberate difference in the categorization of section 3 and section 4 facilitates in the cross-discussion and comparison for the results of closely related research questions. The implications of our results and discussions for different audiences are presented in section 5.

### 4.1 The trends and demographics in security pattern research (RQ1, RQ2, RQ3)

Observing the demographics of security pattern research reveals the continuous research output in the field. The upwards trend in Fig. 4 implies that security patterns are becoming more widely accepted as one of the promising solutions to software security engineering. We identified key researchers and research venues in the field of security patterns to serve as a guideline for those who wish to conduct research in this field. The PLoP and EuroPLoP are currently the premier venues for interested researchers to gather and discuss their findings. The geographical distribution of publications is also helpful in finding geographical hotspots for security pattern research. There is a strong observable correlation between the volume of security patterns research and developed countries. This is because many of the top research institutions and pioneer researchers currently reside in these countries.

The distribution of academia compared to the industry (Fig. 5) is clearly skewed in favor of academia. This signifies that we still know relatively little about the practical settings in which security patterns would flourish. However, the current distribution is not as alarming as it seems when we consider the characteristics of our SMS. We have conducted a review on research papers published in the field. Scholarly publications often stem from academia, where research is the primary objective (as opposed to most industries). If we had instead considered patents and commercial projects regarding security patterns, the distribution in Fig. 5 may be flipped around.

The line graph in Fig. 8 aggregates the trends for the 3 main research topics. Interestingly, the publication trend of the development category closely resembles the overall trend of security pattern publications. This is partly because the development category claims the biggest portion (52%) of the research studies in the field, but it is also a sign of constant development towards new patterns and pattern languages, despite the already large repository of security patterns. For instance, the advent of cloud and mobile computing introduces a new paradigm for security solutions corresponding to the new cloud architectures and mobile security solutions. Therefore, no matter how many solutions have already been proposed as patterns, new security problems in new paradigms will warrant new security patterns and pattern languages. There is an inherent requirement for security patterns to cover newly discovered solutions, but as the solution space of security patterns begins to stabilize, we speculate a shift towards the evolution and refinement of already available patterns and further attempts to better organize and classify patterns.

As seen in Fig. 8, the quality evaluation category has received the least attention. However, as with the other two topics, this topic has also experienced an upwards trend in the recent years. With the increasing number of patterns and pattern languages, and the increased utilization of patterns in software methodologies, quantitative and qualitative evaluation of both the patterns and their impact on software is becoming increasingly crucial. This evaluation can provide insights into how much the current direction of security patterns research will actually improve software security and what alternative directions should be considered. Pattern developers should pay special attention to studies regarding pattern evaluation so as to reduce the number of recurring issues when describing new patterns. Otherwise, new patterns will face the same issues as previous ones and will require unnecessary refinement efforts later on. Researchers working toward pattern usage methodologies should make an effort to address the studies regarding the effect of patterns on software systems and if applicable, refine their methodologies accordingly.

The graph in Fig. 8 also reveals that the pattern usage category is following a similar upwards trend as the pattern development category. This shows the increasing maturity of the field, as the eventual goal of security patterns is to enhance the security of software systems. In order to achieve this goal, the research towards patterns must eventually evolve from isolated solutions to full-fledged software engineering methodologies. The increasing research towards pattern usage can also benefit the evaluation of security patterns, because empirical evaluation requires empirical case studies, which in turn, requires actual usage of security patterns. The current distribution of this category is skewed in favor of research towards integrating security patterns with software development and engineering methodologies to create new approaches or to improve pre-existing methods. Research towards different techniques for selecting and applying security patterns have received less attention. Techniques for selecting appropriate patterns, and applying them at different stages of the development cycle, especially the implementation stage, are an important requirement if using patterns is to be an established practice in industrial software projects.

**4.2 Research efforts towards security pattern development (RQ4, RQ5, RQ6, RQ7)**

Employing standard (and common) notations and templates when describing patterns can provide many benefits. It can facilitate the categorization, evaluation, selection, application, and even teaching of security patterns. While it is understandable that straying away from popular notations and templates can sometimes be necessary, especially when discussing a solution for a particular environment, but not using any standard notation or template should usually be avoided by pattern authors. UML is a well-established notation for modeling the static structure and dynamic behavior of software design. It is also the modeling notation used in every pattern of the GoF pattern catalog, implying its potential for modeling various design solutions. This is the likely reason for its popularity among security pattern authors (69% of studies) UML also provides extensibility mechanisms such as stereotypes, tags, and profiles which might prove useful for modeling security-related attributes. However, these extensibility mechanisms remain largely unexplored by security pattern authors.

Using specific notations such as BMPN and Tropos for describing business or requirements patterns is a plausible alternative to UML, as specific environments might benefit from the specific modeling capabilities of these notations. A rather significant observation is that 27% of papers do not use any standard notation to present their patterns. It is understandable that some patterns describing highly abstract solutions might not benefit much from a standard modeling notation, but in general, avoiding standard notations is a bad practice. Describing patterns using ad hoc and non-standard notations encourages a lack of attention to detail in the structure, behavior, and implementation of the solution. It can also inhibit the effective teaching of patterns to novice designers, which is a commonly mentioned advantage of design patterns. Utilizing standard and common notations can facilitate the evaluation and comparative discussions in the security pattern community. It can also provide a real benefit for pattern application techniques such as model transformation, which can benefit from a common input standard across different patterns (further discussed in section 3.12).

One can assume that the increased fluidity of the POSA template (as opposed to the more rigid structured of the GoF template) is one reason why it is more popular for security patterns (64% of studies), many of which lack the same level of detail as the GoF patterns to properly utilize their template (Bunke, 2015). However, these pattern templates are both quite similar and contain the necessary elements to describe security patterns, which is why some studies use a mixture of elements from both templates. The results in section 3.4 depict that in 22% of the papers, the authors have opted for a custom (often security-centric) pattern template. Security patterns vary from traditional software design patterns based on their intent. Similarly, the users of security patterns will need to categorize, select, and apply security patterns using specific concepts in the field of security engineering. This implies the importance of custom security-centric templates for security patterns. Almost 6% of papers have not used any template to present their patterns. This doesn't necessarily imply an oversight of the authors. Some papers present multiple patterns and also discuss other contributions. This, along with the page limits on many conferences, omit the possibility of presenting patterns in their full standard templates. However, utilizing a standard and shared template to present patterns can facilitate their organization in catalogs and automated selection using data mining (e.g. text mining) techniques. This becomes more apparent as the current collection of security patterns become increasingly larger.

It's promising to see that 83% of papers presenting new patterns have taken care to mention, and sometimes discuss, related patterns. The remaining 17% who don't are often from authors who may not be familiar enough with how patterns should be represented, which is why there is a correlation between papers that do not discuss related patterns and papers that stray from standard notations and templates. As mentioned in section 4.1, describing patterns using ad hoc and non-standard techniques encourages a lack of attention to detail which can allow the deliberate or accidental exclusion of key components in a pattern's presentation. Although, finding related patterns is not a trivial task. Due to the large number of already published patterns, and the variance and incompleteness of most catalogs, it can be hard for authors to investigate every potential pattern that relates to theirs. Systematic studies such as this, are one way to help authors navigate through the different categories of research in the field so as to better identify studies that might include potential related patterns. Establishing a standard (and widely accepted) classification scheme for security patterns can also help in this regard (section 3.7). For instance, if all current patterns are classified into categories based on the boundary points they address (i.e. system core, perimeter, external), future pattern authors can quickly identify which category of patterns have a potential for being related to theirs. Efforts towards creating a pattern language are valuable, because they are a defining factor in the usability of patterns for developers by guiding them through different steps of the development cycle while providing proper pattern-based solutions for each corresponding problem along the way. Simply explaining a solution to an isolated problem without any guidance on the corresponding consequences and next steps will shift developers to look for more holistic approaches in software development.

Security is a spectrum of many different objectives and their corresponding solutions. These range from authentication and authorization to administration and monitoring. Security patterns should also attempt to cover this spectrum, but our findings reveal a very significant imbalance in favor of specific solution categories, with authentication and authorization mechanisms appearing in 36% and 50% of studies. One might argue that this asymmetric nature of the patterns' solution space is not as bad as it seems, as some security objectives might be prioritized higher than others. For instance, a system without a logging mechanism might encounter problems when tracing an imposter's actions in the system, but with proper access control, it can still enforce many of the companies desired security policies and provide a reasonable sense of security. On the other hand, a system without access control will even encounter problems in logging its users' behaviors. Nonetheless, there is already a vast number of studies focused on authentication and authorization patterns (Ciria et al., 2014), (Steinegger et al., 2016), (Fernández et al., 2014), (Priebe et al., 2004) and the field can benefit from more research towards patterns for privacy, monitoring, administration, and recovery. The same can be said about patterns in different types of environments. Table 20 reveals that cloud and distributed systems have considerable patterns while few patterns presented for database systems.

Due to their problem oriented nature, misuse patterns address the limitations of security patterns which are by definition, solution oriented. Misuse patterns first present the problem at hand, such as the threat of information leakage, and then specify the next steps (security patterns) that should be taken, instead of requiring the user to search through a heap of solutions to find the right match. This is a logical approach, as developers are often faced with a collection of problems for which they require corresponding solutions, and not the other way around. These patterns can be used to evaluate the security of existing systems by examining if a given system addresses all of the enumerated threats. They are also valuable for forensic efforts as they identify the assets and relationships exploited before and during the attack, hence serving as a map to find potential traces (Fernandez et al., 2009). Aligning misuse patterns with the corresponding security patterns allows for a deeper discussion regarding the "fitness to purpose" of security patterns against their corresponding threats.

A well-defined and mutually accepted classification of security patterns can benefit both the pattern authors and pattern users in various ways. For instance, if we can achieve an established classification, authors can determine which class their patterns belong to, perhaps by modifying the "related patterns" section in common pattern templates. This can greatly reduce extraneous work for researchers looking to build a pattern language later on. It can also clarify the implicit differences between patterns which might incorrectly be identified as duplicates, while on the other hand, help developers identify and choose between alternative solutions. A good classification scheme should be both complete and mutually exclusive. In other words, it should be capable of classifying all security patterns while ensuring that no pattern can be classified into multiple groups. It should also conform to the taxonomies in the field in order to be suitable for security pattern writers and users. This means that, for example, we should not classify security patterns based on the number of inheritance relationships in their solution's structure. This also implies that some classifications for traditional patterns, such as the one in the GoF book (Gamma et al., 1995), might not be as beneficial to the security pattern community. Security patterns can be viewed as an integration of information security and software engineering. Thus, we should use criteria from both software engineering (e.g. development lifecycle) and information security (e.g. security objective). This itself poses a challenge, as it is hard to prioritize between criteria when attempting to classify patterns. It is also hard to determine whether security patterns should conform to software engineering related or information security related criteria, especially when they are used by a variety of researchers and experts. Therefore, different classifications based on different criteria can benefit different groups of practitioners. However, too much variance in classification strategies hinders standardization, which is a benefit to pattern authors and users alike. Another significant observation is the degree of completeness and mutual exclusivity provided by each criteria, as papers using these criteria do not claim to have collected every pattern for classification. A study by Hafiz et al. (Hafiz et al., 2007) reveals the deficiencies in many of the criteria mentioned in Table 19 and proposes a new hierarchical classification scheme based on multiple criteria. This hierarchical scheme is especially effective at classifying security patterns at different levels of abstraction, while ensuring that no two patterns at a single level of the hierarchy can be classified into multiple groups. A more recent work by Ponde et al. (Ponde et al., 2016) uses unsupervised hierarchical clustering to extract related groups of security patterns using a set of dominant criteria. This systematic mapping study has shown that the rapid expansion of security patterns has produced a large body of work. Classifying patterns can greatly help the applicability of security patterns by allowing developer to quickly find the solution to their problem. It can also help pattern authors to better express how their pattern fits into the overall security pattern landscape.

### 4.3 Research efforts towards security pattern evaluation (RQ8, RQ9)

The observations in the results of RQ8 reveal that the vast majority (71%) of evaluations use qualitative techniques. This implies the need for more quantitative evaluations in the field, which has also been expressed by others, such as Halkidis et al. (Halkidis et al., 2006). Quantitative evaluation methods have many benefits. They are more mathematically precise, less prone to bias, automatable, and easier to verify and replicate. However, quantitative evaluation relies on quantifying often complex concepts. This can result in a significant loss of information, which in turn, creates a discrepancy between "what we should be evaluating" and "what we are actually evaluating". This reveals the lesser discussed advantage of qualitative evaluation, i.e. the ability to provide a more holistic and meaningful assessment. Nonetheless, qualitative techniques often rely on experts' opinion and case studies to evaluate patterns, and they are therefore more prone to subjective opinions or bias.

As mentioned earlier, quantitative evaluation retains several benefits, but researchers should take into account the considerations regarding its objectivity. For instance, studies which use the help of experts to assess the quality of security patterns should make an effort to include a variety of experts from diverse backgrounds, as experts in academia and the industry can have different priorities in their evaluation criteria. Even diverse experts in the industry, such as web service developers versus middleware developers, can have varying opinions. Of course, collecting an inclusive group of experts from all background may not be feasible, but ideally, they should be comparable to the audience which are targeted by security pattern research. Studies that use empirical studies should try to utilize control and treatment groups which are adequately representative of software developers who would benefit from security patterns. From the empirical studies we observed, many of them use undergraduate and graduate computer science students to conduct their experiments (Abramov et al., 2012b), (Yskout et al., 2015), (Smith and Williams, 2012). While students are an accessible resource in academic research, their motivation and expertise might not always be representative of software developers.

Evaluating software systems before the implementation stage is a difficult task, as design decisions can be implemented using various techniques, and this can greatly influence the final product. Although evaluating software at the design stage is not easy, evaluating patterns at the design stage is even more challenging. This is due to the fact that patterns should provide generalized and fairly abstract solutions which can be applicable to a wide range of concrete problems with similar characteristics. In other words, a class diagram for a design pattern contains less detailed information than a class diagram for a real software project. Attempts to evaluate security at the design stage have implied the need to enrich the class diagram beforehand (Alshammari et al., 2009), (Alshammari et al., 2010). This signifies the need to go beyond the simple (non-extended) UML class diagrams to present security patterns. It is also worth noting that many security patterns might even address more abstract stages of the development lifecycle e.g. architecture design and requirements analysis, which can be even harder to evaluate. Nonetheless, evaluating the quality of security patterns and the effect they have on the target software is an important task. This evaluation can provide insights into how much the current direction of security patterns research will actually improve software security and whether alternative directions should be considered

The observations in the result of RQ9 reveal that 39% of studies have conducted pattern-specific evaluations. These types of "bounded" evaluations can seem at a disadvantage as their methods and results may not be applicable to patterns outside the scope of that paper. However, there are certainly a number of benefits for these kinds of evaluations. First of all, proving an evaluation technique to be applicable to every pattern in every scenario is a near-impossible task, and attempting to assess a diverse set of patterns in diverse environments forces researchers to generalize the pattern-specific and target-specific features in an effort to pursue a holistic evaluation technique. This hinders the usefulness of the evaluation and its results. Bounded evaluations allow the authors to focus closely on a set of concrete patterns, and better understand their intent and target systems, which is important when evaluating patterns. It is also a means to focus evaluation efforts towards more important and commonly used patterns in the research community.

Our observations in section 3.9 reveal that the majority of studies in the evaluation category are convinced of the usefulness of security patterns for creating more secure designs and reducing the number of undesirable design alternatives in the process. Others are mostly concerned about some of the current shortcomings such as the lack of detail in pattern specifications which impede the proper utilization of patterns. These shortcomings should not be ignored by security pattern researchers which strive for the increased maturity of the field.

**4.4 Research efforts towards security pattern usage (RQ10, RQ11, RQ12)**

The most prominent objective of security patterns is to improve the security of software during development. In order to achieve this goal, the research towards patterns must eventually evolve from isolated solutions to full-fledged software engineering methodologies. Using patterns in different software environments might require specific considerations, therefore, a significant portion of research towards pattern usage is aimed at specific environments. The variety in usage environments also depicts the wide range of applications for security patterns. Distributed and cloud systems add up to 32% of studies bounded to specific environments. This indicates the awareness of the security pattern research community in regards to the current software trends towards distributed and cloud computing.

Discussing patterns in the context of specific environments provides the possibility to analyze the applicability of security patterns in their current form, to emerging paradigms such as cloud and embedded systems, and determines whether the current pattern presentation formats are adequate for these relatively new settings. This can provide guidelines to direct further development efforts to cater to the needs of emerging trends in computing, because as mentioned in section 3.4, the majority of the current security patterns are presented using traditional pattern templates such as the GoF (Gamma et al., 1995), which were first introduced before the emergence of new paradigms such as cloud computing. More research towards pattern usage can also prove beneficial in the evaluation of security patterns, as empirical evaluation requires empirical case studies, which in turn, requires the actual usage of security patterns. All of this signals an important benefit for research in the pattern usage category, namely, analyzing the suitability of current approaches in the pattern development and pattern evaluation category.

Comparing the results of the specific environments for applying patterns (RQ10) and the results of the specific patterns used in different papers (RQ11) reveals a logical correlation. Often, but not always, studies which have bounded their scope to a specific environment (section 3.10) have also used a specific set of patterns. For example, Hamid et al. (Hamid et al., 2013) have used S&D patterns because their research is focused on embedded systems. Another example is the work of Fernandez and Monge (Fernandez and Monge, 2014) in which they use cloud-specific patterns when discussing a security reference architecture for cloud systems. Attempting to research the usage of security patterns by limiting the scope to a specific environment or a specific set of patterns might have some drawbacks. It may be that the mentioned set of patterns in the specific environment does not adequately represent the type of usage that pattern users are interested in, which might raise concerns about the applicability of the results. However, attempting to present an approach to identically utilize patterns in multiple diverse scenarios pressures researchers into overly general approaches that can impede the usefulness of patterns for software projects, which are often developed for specific needs in specific environments. Focusing pattern usage research towards a specific group of patterns allows for deeper analysis and more specific selection and application methods. Nonetheless, researchers should aim to align the environment-specific methodologies for using patterns in the software development lifecycle with the specific patterns that corresponds to those environments.

The majority of the studies we have observed utilize model transformation techniques to apply patterns in the design or code. This lack of variety might be due to the inherent nature of pattern application, where we possess a model as the input and we strive for an extended model as the output (i.e. model transformation). In any case, research towards concrete techniques for pattern selection and application can be used as a measure for the feasibility of current pattern-based methodologies. When devising techniques for selecting and applying security patterns, exploring the applicability of the corresponding techniques in traditional design patterns can prove beneficial. Although security design patterns and traditional design patterns vary in their intent, as pointed out by section 3.4, the notations for explaining their structural and behavioral aspects are quite similar. Although some selection techniques such as ontology, classification, and feature models may be specific to each group of patterns, the main concepts involved in these techniques can be used across different fields. Also, many application techniques utilize model transformation, which transforms the solution model of the pattern into a usable construct for software development (e.g. code). These transformation rules differ based on the input and output structure, rather than the intent of the input or output. Therefore, model transformation techniques for design patterns can also be used for security patterns. This can prove beneficial as the research in pattern application is more mature in the field of design patterns, which explores application techniques as part of larger refactoring procedures (Tsantalis and Chatzigeorgiou, 2010), (Zafeiris et al., 2017).

## 5 IMPLICATIONS OF THE FINDINGS

We have conducted a systematic review of security pattern research in order to present a guideline for researchers, practitioners, and teachers interested in this field. Due to the breadth and depth of this review, different groups of audiences will find significant implications implied in our results and discussions (section 3 and 4). Each of the results in previous sections may have implications for multiple audiences. This section presents the implications of this SMS for interested researchers, practitioners, and teachers. The implications for each group include the various ways in which they can benefit from the results of this study and how the results can guide the path to improving the current state of the art.

**Implications for researchers**
1. There is a rather large difference between the amount of security pattern research in academia compared to the industry (RQ1). As we discussed in section 4.1, this may not be as alarming as it seems due to the nature of our systematic mapping study which focuses on research papers (commonly attributed to the academic domain). Nonetheless, it signifies the shortage of evidence for the practical settings in which security patterns have been (or can be) successful. It would be interesting to see how this distribution would change if we focused on software projects and patents related to security patterns. This presents a potential path for future empirical studies which should focus on the maturity and popularity of security patterns in active software projects and patents.
2. Security pattern researchers have opted for the popular UML notation to present their patterns (RQ4). However, many of the favored UML extensibility mechanisms such as stereotypes, tags, and profiles remain largely unexplored by the pattern authors. These mechanisms can prove useful in modeling security-related attributes and behavior. It can also facilitate quantitative evaluation using security design metrics which rely on enriched UML notations such as UMLSec (Jürjens, 2005) as discussed in section 4.3. A similar argument can be made for using a custom pattern specification template which incorporates security-centric guidance and the support for emerging paradigms (such as cloud computing), as opposed to the GoF and POSA templates used for traditional design patterns.

3. In order for security patterns to be accepted as a holistic solution to secure software development, researchers should pay special attention to the relationship between patterns. This can help in guiding developers along a step-by-step process to applying patterns specific to the problem at hand. It is also a means to differentiate between the benefits and drawbacks of each design alternative. Current security pattern catalogs are incomplete, and disagree on classification criteria (RQ7). Therefore, a timely effort towards a comprehensive, widely-accepted, and up-to-date security pattern catalog can greatly benefit the research community (RQ5). This is an iterative process, as the relationships between the patterns in the catalog can be extracted from the *related patterns* mentioned in each pattern's specification, but for the authors to specify which patterns relate to theirs, they require a categorized catalog of patterns. Due to the continuous growth of research towards new patterns (RQ2), a lack of a comprehensive and well-established catalog will result in a large, unstructured, and confusing solution space for practitioners.
4. Among the 274 papers reviewed for this study, only 8% have devoted their efforts towards evaluation, whereas the development and usage of security patterns are respectively responsible for 52% and 35% of the research (RQ2). Continuous efforts towards developing new patterns and their usage, without the accompanying research required to evaluate their effect, can cause major problems later on. Evaluating the different facets of security pattern research can help identify and prevent incorrect, inapplicable, and redundant approaches chosen by current researchers in the field. In order to prioritize efforts, future evaluations can focus on security patterns which correspond to commonly used security solutions in the industry.
5. The majority of the evaluation efforts rely on qualitative approaches (RQ8) which are believed to be less precise and more prone to subjectivity as opposed to quantitative methods. Formally specifying security patterns is one possible solution for evaluating existing systems, but formal specifications are too precise and against the spirit of patterns which should be general and relatively abstract solutions. Another is the definition of evaluation metrics which can assess the quality of individual patterns, or a collection of patterns used in the design. As mentioned earlier, this might require enriched UML notations and security-centric templates when specifying security patterns. The consensus of the current qualitative evaluations on security patterns maintain a favorable position towards security patterns. However, many of these evaluations rely on empirical data when exposing different case studies to a group of experts, mostly in academia. Future studies should also make an effort to include a variety of experts from diverse backgrounds, as experts in academia and the industry can have different priorities in their evaluation criteria. Although most evaluation studies are optimistic, some evaluations have expressed their concerns about the usefulness of security patterns in their current form. These studies are mostly concerned about some of the current shortcomings such as the lack of detail in pattern specifications which disrupt the proper use of patterns. These issues should not be ignored by security pattern researchers as they directly impact the maturity of the field.
6. There is a large volume of research concerning security pattern usage, many of which cater to specific environments and patterns (RQ10 and RQ11). This signifies the increasing maturity in the field and the awareness of the security pattern community with respect to emerging paradigms such as distributed and cloud computing. There are many efforts towards software development methodologies using security patterns, pattern selection approaches, and pattern application techniques. However, these efforts are not currently aggregated into a holistic and tool-supported approach. Software developers should be able to rely on a unified process which allows them to search for the appropriate patterns based on their security problem, analyze the benefits and drawbacks of different patterns, and apply the desired choice into their design and implementation. If this vision is to be realized, it requires more than just research in the *usage* category, as we need pattern development efforts to improve implementation details and pattern evaluation efforts to assess the applicability of the pattern selection and application techniques. It can be worthwhile to draw on the experiences of pattern application in the more mature field of software design patterns (non-security), which explores application techniques as part of larger refactoring procedures.

**Implications for practitioners**
1. Security patterns are becoming one of the well-established approaches to design and develop secure software. There is an abundance of research in different aspects of security patterns which is gradually matured through 21 years of research experience (RQ1). Furthermore, the patterns themselves are extracted from commonly used solutions which have proved suitable in past development experiences. Security practitioners should utilize this knowledge to improve their security designs and avoid less-than-optimal alternative solutions.
2. Current research in developing and using security patterns have been mostly deprived of industry experience (RQ1). Security practitioners should also play a major role in improving the current state of the art. Industry practitioners should take part in premier conferences in the field (RQ3) and provide their perspective on the usefulness of current and future research approaches with respect to the industry. This includes views on how security patterns should be presented (RQ4 and RQ5) and how they should be categorized (RQ7), as well as computing environments which would benefit more from pattern-based solutions (RQ10).
3. Security patterns research is currently ripe for empirical evaluations (RQ1 and RQ8). There is a current scarcity of empirical evidence from the industry in what metrics should be used in empirical evaluations. Industry experts should routinely collaborate with researchers as to provide insights into appropriate and industry-relevant metrics for empirically evaluating the effect of security patterns on developers with different levels of technical expertise. Industry practitioners should also collaborate with researchers to report their successful and unsuccessful experiences with patterns in different types of software projects. These collaborations will help the practitioners by bringing security patterns even closer to what is for successful industry applications.
4. Misuse patterns are an emerging concept which has multiple implications for practitioners (RQ6). Due to the problem oriented nature of these patterns, practitioners can pair and use misuse patterns with security patterns. This is an advantage for developer as they are often faced with a set of problems for which they require corresponding solutions, and not the other way around. Misuse patterns can also prove useful in forensic activities by providing a checklist of all the components affected during the attack. These patterns identify the assets and relationships exploited before and during the attack, and can serve as a map for finding potential traces

**Implications for teachers**
1. With over 21 years of experience, security patterns are becoming one of the promising solutions to software security engineering (RQ1). These patterns embody expert knowledge and best practices. Therefore, a commonly stated benefit of patterns is the dissemination of expert knowledge and experience to novice developers. Teachers of information security and software engineering, whether in university courses or professional courses, should incorporate security patterns into their syllabus. This can result in a faster achievement of suitably secure designs by guiding novices to make similar decisions as the experts (RQ9). The vast amount of security pattern research for specific environments (RQ10 and RQ11) can also benefit instructors who are teaching focused information security courses such as "database security" and "security in cloud architectures". They can also utilize misuse pattern to augment "forensics" security courses (RQ6).
2. Teaching expert knowledge to novice designers is one of the main objectives of security patterns. Therefore, information security instructors should play a major role in helping the research community move forward. By taking part in premier conferences and working closely with pioneer researchers (RQ3), teachers can discuss the teaching requirements for security patterns with other members of the community. Helping students better understand the expert knowledge encompassed in these patterns should be one of the driving factors in choosing a template for presenting patterns, design notation for modeling pattern structure and behavior, and pattern classification criteria (RQ4 and RQ7). Teachers should also report their student's positive and negative experiences with patterns in order to facilitate empirical evaluation (RQ8 and RQ9).

## 6 CONCLUSIONS AND FUTURE WORK

We have conducted a systematic mapping study in the field of security patterns from 1997 to the end of 2017. We retrieved a total of 403 relevant papers and used 274 of them (based on quality criteria) for our analysis. We have spent a significant portion of the review on devising a suitable research methodology and presented our final methodology in this paper. This includes a comprehensive 3-tier search strategy comprised of manual search, backward snowballing, and database search on 4 scientific libraries, followed by an evaluation of our search which yielded a more than acceptable result of 95.5%. We have also included a comprehensive supplementary document to present the details of our systematic study which were too large to fit in this paper.

We have defined a set of 12 research questions for our systematic mapping study. The first 3 research questions address the demographics of security pattern research such as topic classification, trends, and distribution between academia and the industry, along with prominent researchers and venues. The next 9 research questions focus on more in-depth analyses such as pattern presentation notations and classification criteria, pattern evaluation techniques, and pattern usage environments. After answering the research questions and discussing their results, we gathered the implications of our findings for researchers, practitioner, and teachers in the information security and software engineering community.

Our investigations reveal that *security patterns* is an active and growing field of research across various geographical locations. The increased development of patterns corresponds to new software engineering paradigms (for instance the advent of cloud computing) which require new security solutions, and the increased usage of patterns is a sign of increased maturity in the field. Security patterns are increasingly being used to develop new software development methodologies or improve existing approaches. Systematic mapping studies such as this can be used as a basis for more specific systematic literature reviews. In future works, each of the classified topics in our research tree can be further investigated to answer more specific research questions. The implications for researchers, practitioners, and teachers is also a guideline on how different audiences can take advantage of current research and also take part in helping the field towards further maturity.

## ACKNOWLEDGEMENTS

The authors would like to thank members of the Software Quality Lab at Ferdowsi University of Mashhad, especially Bahareh B. Mayvan and Ali A. Ramaki, for their support throughout this systematic mapping study.